\documentclass[12pt,preprint]{aastex}
 \newcommand{\beq}{\begin{equation}}
 \newcommand{\eeq}{\end{equation}}
 \newcommand{\beqn}{\begin{eqnarray}}
 \newcommand{\eeqn}{\end{eqnarray}}

\begin{document}
\title{
A Gaia based photometric and kinematic analysis of the old open cluster King 11}
\author{Devesh P. Sariya$^1$,
Ing-Guey Jiang$^1$,
D. Bisht$^2$,
M. D. Sizova$^3$, 
N. V. Chupina$^3$, 
S.V. Vereshchagin$^3$,
R. K. S. Yadav$^4$,
G. Rangwal$^5$
}
\affil{
{$^1$Department of Physics and Institute of Astronomy,}\\
{National Tsing-Hua University, Hsin-Chu, Taiwan}\\
{$^2$Key Laboratory for Researches in Galaxies and Cosmology,}\\ 
{University of Science and Technology of China, Chinese
Academy of Sciences, Hefei, Anhui, 230026, China}\\
{$^3$Institute of Astronomy Russian Academy of Sciences (INASAN),}\\ 
{48 Pyatnitskaya st., Moscow, Russia}\\
{$^4$Aryabhatta Research Institute of Observational Sciences,}\\ 
{Manora Peak, Nainital 263002, India}\\
{$^5$Center of Advanced Study, Department of Physics, D. S. B. Campus,}\\ 
{Kumaun University Nainital 263002, India}
}

\begin{abstract}

This paper presents an investigation of an old age open cluster King 11 
using Gaia's Early Data Release 3 (EDR3) data. 
Considering the stars with membership probability ($P_{\mu}$) $> 90\%$,
we identified 676 most probable cluster members within the cluster's 
limiting radius. 
The mean proper motion (PM) for King 11 is determined as: 
$\mu_{x}=-3.391\pm0.006$ and $\mu_{y}=-0.660\pm0.004$ mas yr$^{-1}$.
The blue straggler stars (BSS) of King 11
show a centrally concentrated radial distribution. 
The values of limiting radius, age, 
and distance are determined as
18.51 arcmin, 3.63$\pm$0.42 Gyr and $3.33\pm0.15$ kpc, respectively. 
The cluster's apex coordinates
($A=267.84^{\circ} \pm 1.01^{\circ}$, $D=-27.48^{\circ} \pm 1.03^{\circ}$)
are determined 
using the apex diagram (AD) method 
and verified using the ($\mu_U$,$\mu_T$) diagram. 
We also obtained the orbit that the cluster follows in the Galaxy
and estimated its tentative birthplace in the disk.
The resulting spatial velocity of King 11 is 60.2 $\pm$ 2.16 km s$^{-1}$. 
A significant oscillation along the $Z$-coordinate up to 
0.556$\pm$0.022~kpc is determined.

\end{abstract}


\section{Introduction}
\label{INTRO}

Open clusters are the relatively younger and metal-rich
(as compared to the globular clusters) star clusters that occupy the 
space in the Galactic plane. They are important tools to study
the formation and evolution history of the Galactic disc
(Chen et al. 2003; Moraux 2016).

Proper motion (PM) is the displacement 
in the position of a celestial object with time. 
It is associated with astrometry which deals with the determination of 
positions of the objects in the sky.
By virtue, astrometry can be considered the oldest branch of astronomy.
The derivation of the PMs has been a tedious job 
and it used to require data spanning several decades 
to derive PMs with an accuracy of $\sim$1 mas yr$^{-1}$ (Cudworth 1986). 
The situation was improved by the the usage of the 
charge-coupled device (CCD) data. 
The space mission Hipparcos (ESA 1997) provided PMs
and parallax values for the brighter stars.
In the case of star clusters, 
the fainter main sequence is usually indistinguishable from the field stars
(Jones 1997; Piotto et al. 2004). 
This hampers the possibility of correctly defining the cluster parameters
and mass function slope. 
Hence, the cluster membership status of the stars becomes pivotal.
The selected cluster members would be useful not only for the photometric 
and kinematical studies, but also for selecting the
candidates for spectroscopic observations (Cudworth 1997; Anderson et al. 2006). 
Hubble Space Telescope (HST) provided PMs up to deeper magnitudes
(Bedin et al. 2001; King \& Anderson 2002).
Also, efforts were made to utilize the ground based wide-field CCD imagers
to study cluster PMs 
(Anderson et al. 2006; Yadav et al. 2008, 2013; 
Bellini et al. 2009; Sariya et al. 2017, 2018).
Still, there was a need of the space based satellite specially designed and 
devoted to the astrometry. 
Gaia (Gaia Collaboration et al. 2016 a,b)
has unlocked the doors of many scientific opportunities.
For the star clusters, several important studies are done 
using Gaia data (e.g. Cantat-Gaudin et al. 2018; Soubiran et al. 2018; 
Postnikova et al. 2020; Bisht et al. 2020, 2021a,b; Ferreira et al. 2021;
Garro et al. 2021; Rain et al. 2021; Sariya et al. 2021a,b
Shull et al. 2021; Xiaoying et al. 2021). 

Gaia data can also be utilized in the kinematical study of the clusters.
Using the astrometric solution from Gaia, 
we can determine the apex point of the clusters
(Elsanhoury et al. 2018; Postnikova et al. 2020; 
Bisht et al. 2020; Sariya et al. 2021a).
Gaia's astrometric parameters, combined with radial velocities, 
make it possible to trace the motion of a star cluster in the Galaxy 
and determine the tentative place of its birth. 
This analysis is crucial to navigate the evolution of the Galaxy.

King 11 ($\alpha_{2000} = 23^{h}47^{m}38.88^{s}$, 
$\delta_{2000}=+68^{\circ} 38^{\prime} 09.6^{\prime\prime}$;
$l$=117$^\circ$.151, 
$b$=6$^\circ$.484 Cantat-Gaudin et al. 2018)
is an old open cluster lying in the second Galactic quadrant.
The cluster has a high reddening as  
$E(B-V)$ is reported in the range 0.90--1.06 by various authors 
(Aparicio et al. 1991, Friel et al. 2002, Tosi et al. 2007,
Kyeong et al. 2011, Kharchenko et al. 2013).
Several photometric studies of this cluster are available in literature
(Kaluzny 1989; Aparicio et al 1991; Phelps et al. 1994; Tosi et al. 2007;
Kyeong et al. 2011). 
Scott et al. (1995) provided mean radial velocity of the cluster
($-35\pm16$ km s$^{-1}$).
The age of the cluster varies in the literature.
According to Salaris et al. (2004), King 11 should be 5.5 Gyr old,
while Tosi et al. (2007) quote an age range of 3.5--4.75 Gyr. 
The PM and distance of the cluster is given as:
($-$3.358, $-$0.643) mas yr$^{-1}$ 
and 3433.2 pc by Cantat-Gaudin et al. (2018).

The data used in the current study are discussed in Section~\ref{OBS}.
Based on the PMs, initial decontamination of the field stars
is shown in Section~\ref{PM}, where we also determine
the membership probability for the stars.
The most probable cluster members are then taken to 
study the structural and fundamental parameters of King 11 in Section~\ref{FUNDA}.
The apex coordinates are derived in Section~\ref{adapex}.
Section~\ref{orbit} presents an analysis and discussion 
on the orbit of King 11 in the Galaxy. 
The outcomes of the present study are summarized in Section~\ref{CON}.

\section{Data used}
\label{OBS}

%%%%%%%%%%%%%%%%
\begin{figure*}
\begin{center}
\includegraphics[width=9.5cm, height=9.5cm]{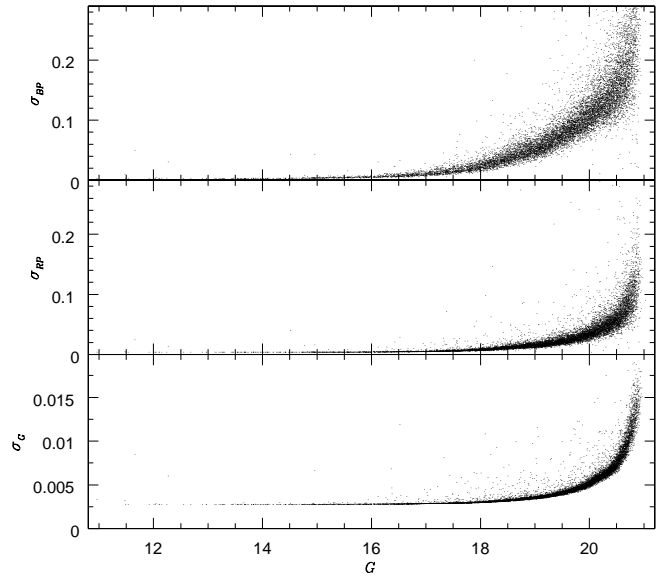}
\caption{Photometric errors in Gaia bands ($G$, $BP$ and $RP$) with $G$ magnitudes.}
\label{error_mag}
\end{center}
\end{figure*}
%%%%%%%%%%%%%%%%

\begin{figure*}
\begin{center}
\includegraphics[width=9.5cm, height=9.5cm]{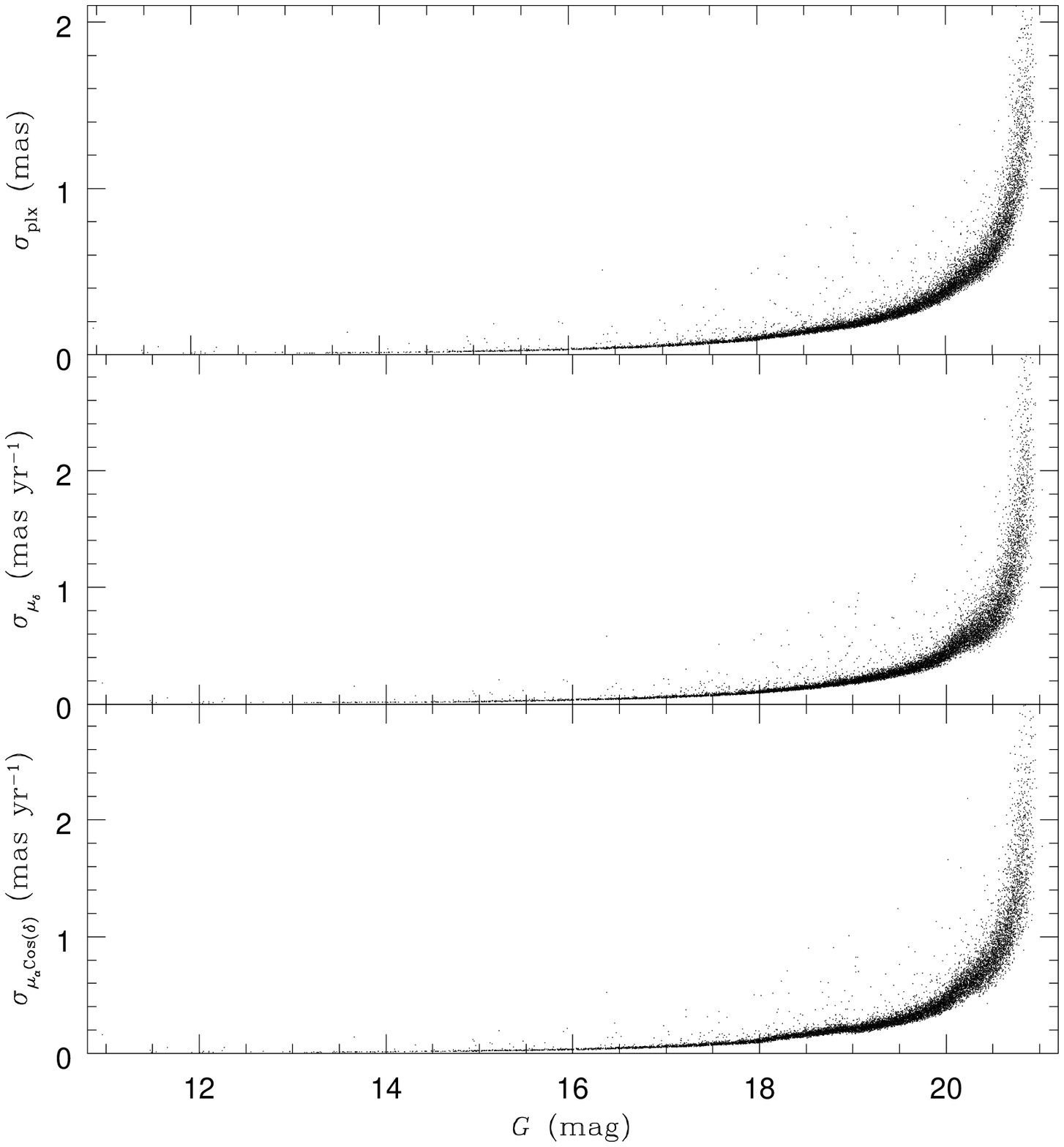}
\caption{Plot of parallax errors (top panel) and 
PM errors (middle and bottom panels) versus $G$ magnitude.}
\label{error_pm}
\end{center}
\end{figure*}

We used Gaia-EDR3 data 
(Gaia Collaboration et al. 2020) for the study of King 11.
Gaia-EDR3 data provides five parameter astrometric solution i.e.
positions ($\alpha, \delta$), parallaxes 
and PMs ($\mu_{\alpha} cos\delta , \mu_{\delta}$).
The radial velocities are also available in Gaia-EDR3 for relatively less number of stars.
The photometry from Gaia is available in three pass bands, 
namely, the white-light $G$, the blue $BP$ and the red $RP$ bands. The photometric
errors in three bands versus $G$ mag is shown in Fig \ref{error_mag}. 
The errors in parallax and PMs are shown in Fig.~\ref{error_pm}.
The mean PM errors for the data we used is
$\sim0.04$~mas~yr$^{-1}$ for $G<$17 mag which increases up to
$\sim0.2$~mas~yr$^{-1}$ for $G<$20 mag.
For stars brighter than 20 $G$ mag, the mean error in parallax is $\sim0.2$~mas.
For $G<$20 mag, the median values of photometric errors
in $G, BP$ and $RP$ are 0.003, 0.053 and 0.015 mag respectively.

%%%%%%%%%%%%%%%%
%%%%%%%%%%%%%%%%

\section{Cluster membership}
\label{PM}

\subsection{Vector point diagrams}
\label{VPD}

In a cluster's region, the precise PMs from Gaia can be adequate enough
to provide a preliminary selection of cluster members.
In a plot between both the PM components
($\mu_{\alpha} cos{\delta}$, $\mu_{\delta}$), 
known as the vector point diagram (VPD), 
the distribution of the cluster stars is centered around a common point.

\begin{figure*}
\begin{center}
\includegraphics[width=15.5cm, height=15.5cm]{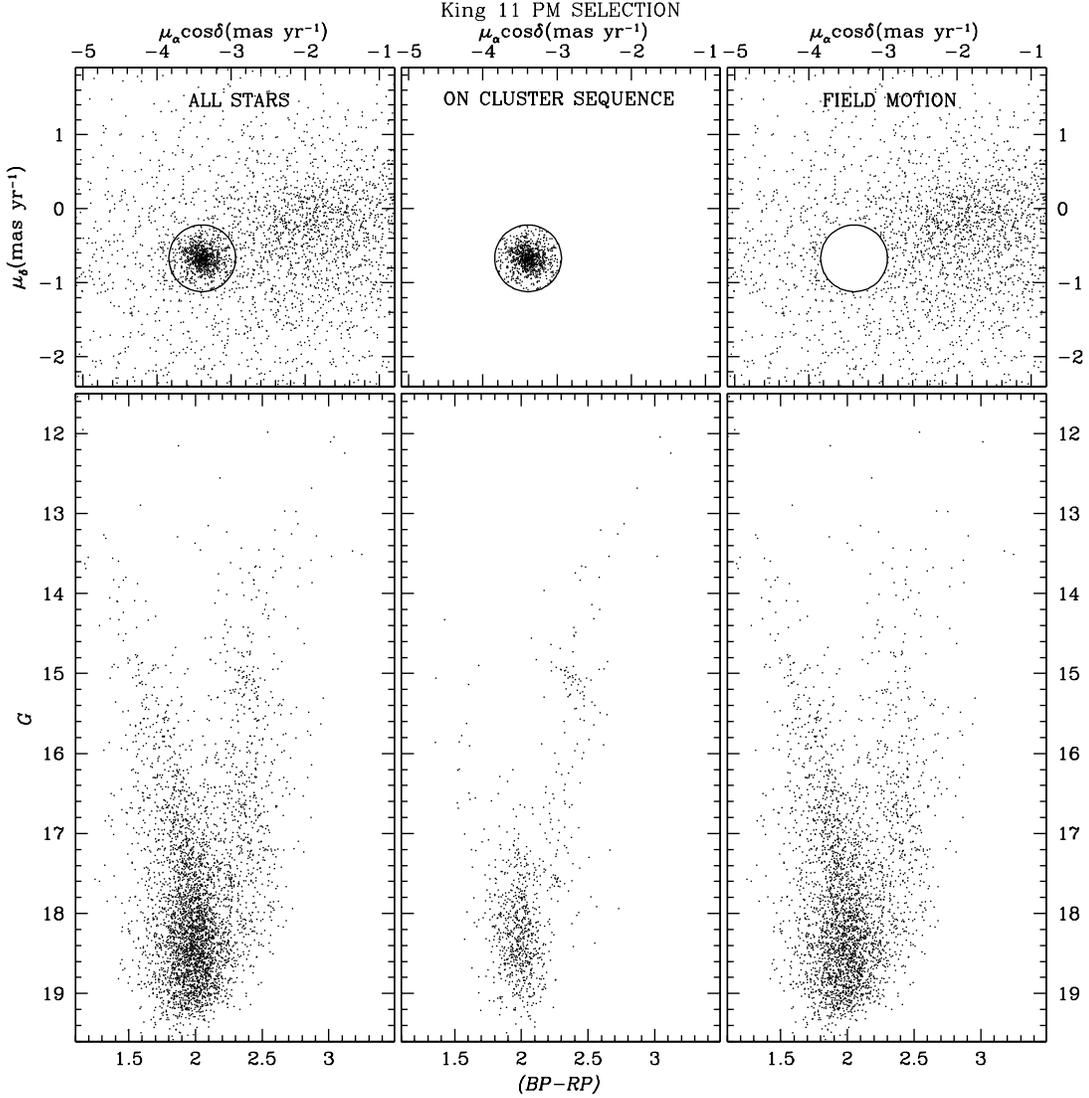}
\caption{(Top panels) The PM VPDs for the stars in the direction of King 11. 
(Bottom panels) $G$ versus $(BP-RP)$ CMDs. 
(Left) The entire sample. 
(Center) Stars lying within the circle of radius of 0.45~ mas~ yr$^{-1}$ in the VPD.
(Right) The background/foreground stars in the direction of the cluster.
Here, we have used only the stars with PM errors $<$ 1~mas~ yr$^{-1}$
and parallax errors $<$ 1 mas.}
\label{vpd}
\end{center}
\end{figure*}

For King 11, the VPDs are shown in the top panels of Fig.~\ref{vpd}.
The bottom panels of the figure present $G$ versus $(BP-RP)$ 
color-magnitude diagrams (CMDs). 
All the stars plotted in Fig.~\ref{vpd} have PM errors
less than 1 mas yr$^{-1}$ and parallax errors better than 1 mas.
Going from the left to the right panels  in this figure, 
we show: the entire sample of stars, the preliminary cluster members,
and population of the non-member stars.
The non-member stars lie in the field-of-view of the cluster, 
either in the background or foreground
of the cluster.
In the VPDs, a circle is shown, which exhibits the fact that
the cluster's member stars
follow the same mean PM (Sariya et al. 2021b).
The radius of the circle is 0.45 mas yr$^{-1}$
which is chosen after trying various radii 
and judging their effect on the CMDs shown in the bottom panels. 
This cluster has a high reddening value 
($E(BP-RP)$ = 1.19) as calculated in Section~\ref{ISOCHRONE}
using the visual fitting of theoretical isochrones on the CMD.
Due to the high reddening, 
the cluster's sequences are not sharply defined in the middle CMD
(the presumed cluster members).

%%%%%%%%%%%%%%%%
%%%%%%%%%%%%%%%%

%%%%%%%%%%%%%%%%
\begin{figure*}
\begin{center}
\hbox{
\includegraphics[width=8.5cm, height=8.5cm]{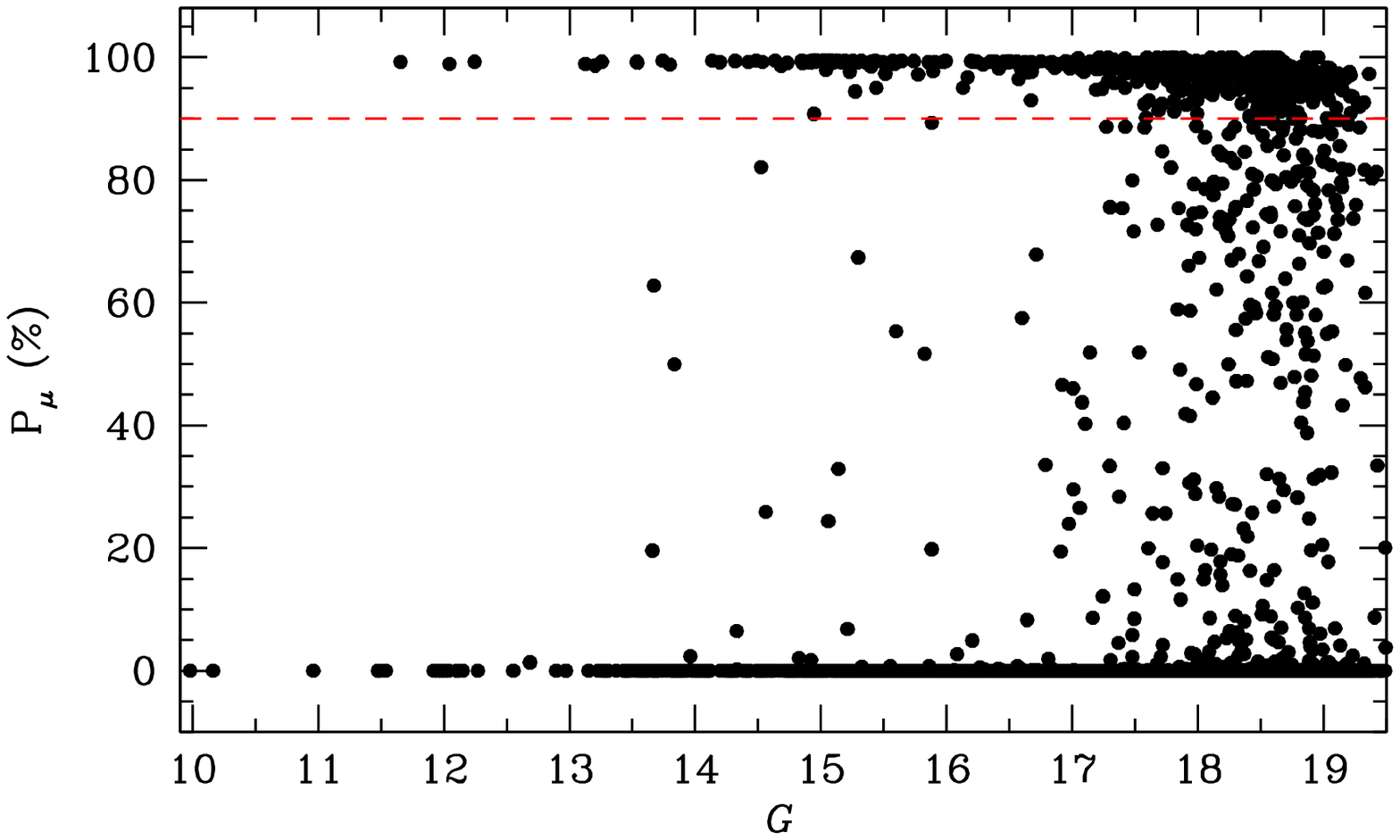}
\includegraphics[width=8.5cm, height=8.5cm]{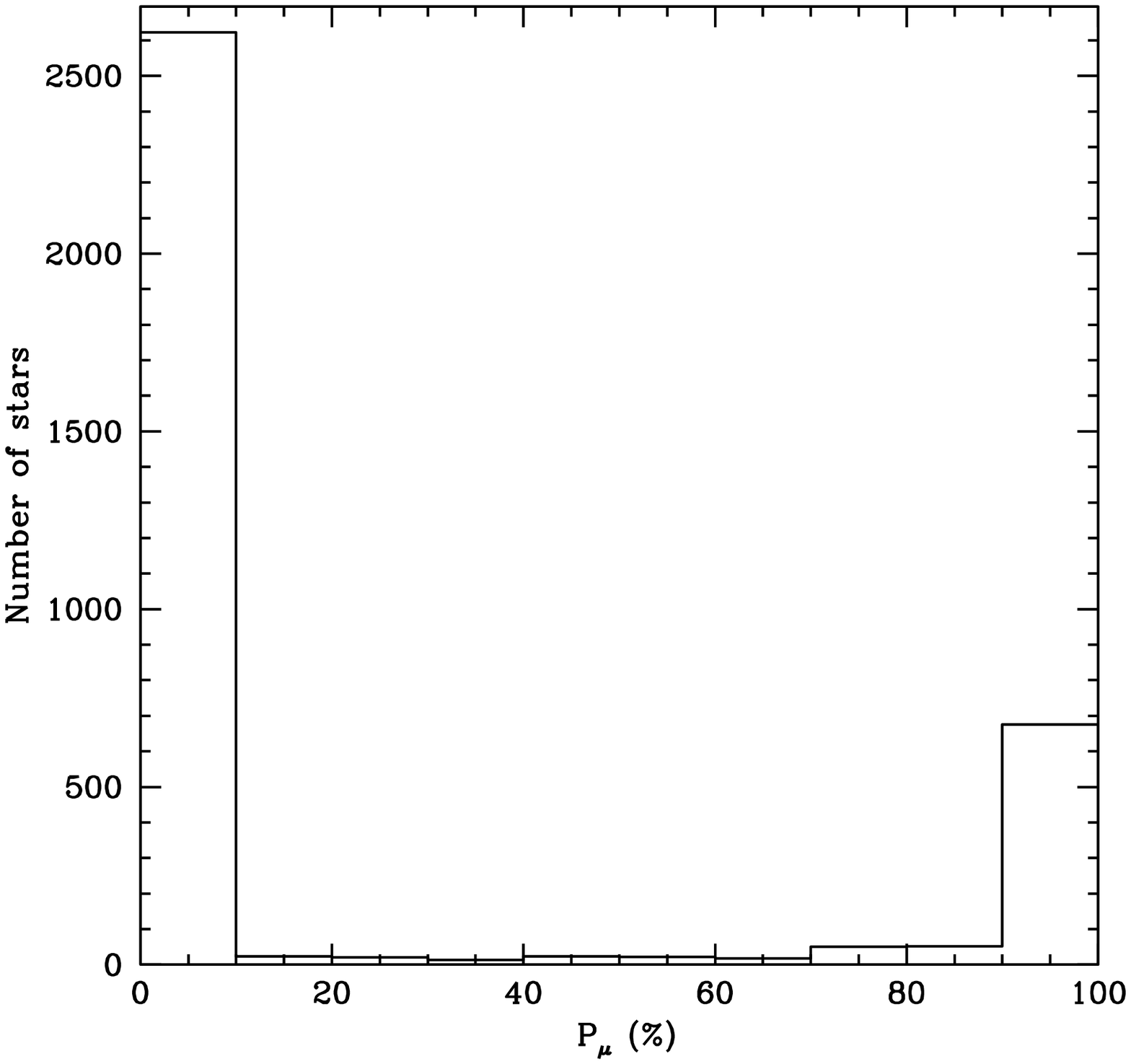}
}
\caption{(left panel) Distribution of the membership probability of stars as a function of $G$
magnitude. We have considered probable members having membership probability higher than 90$\%$ as
indicated by horizontal dash line. (right panel) Membership probability histogram for all stars in
the area of King 11.}
\label{mp_hist}
\end{center}
\end{figure*}
%%%%%%%%%%%%%%%%
%%%%%%%%%%%%%%%%%%%%%%%%%%%%%%%%%%%%%%%%%%%%%%%%%%%%%%%%%%%%%%%%%%%%%%%%%%
\begin{figure*}
\begin{center}
\includegraphics[width=8cm, height=8cm]{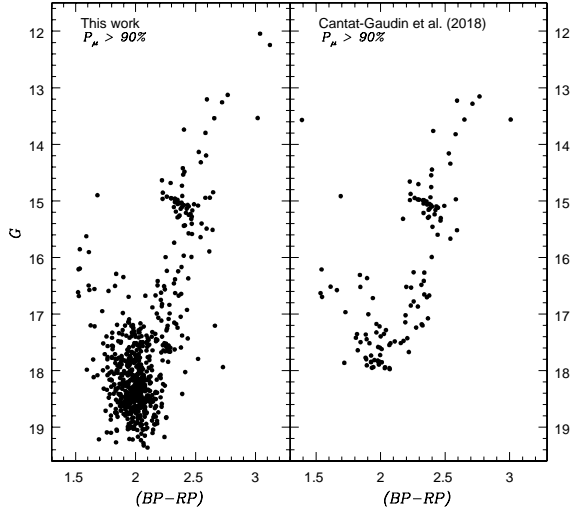}
\caption{CMD of the most probable member stars of King 11 shown from this work (left panel)
and from Cantat-Gaudin et al. (2018) in the right panel.} 
\label{figmp90}
\end{center}
\end{figure*}
%%%%%%%%%%%%%%%%%%%%%%%%%%%%%%%%%%%%%%%%%%%%%%%%%%%%%%%%%%%%%%%%%%%%%%%%%%%

%%%%%%%%%%%%%%%%
\begin{figure*}
\begin{center}
\includegraphics[width=8.5cm, height=8.5cm]{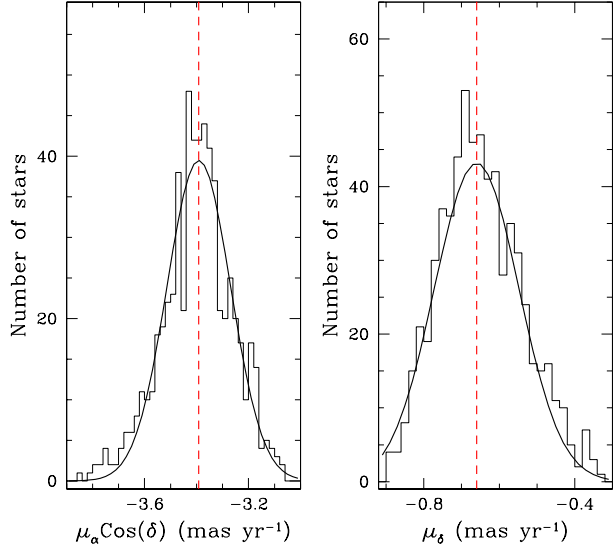}
\caption{Histogram for the determination of mean values of PM 
in RA and DEC directions. The Gaussian function fitted to 
the central bins provide the mean values. 
The vertical dashed lines in both panels indicate the calculated 
mean values.
}
\label{pm_hist}
\end{center}
\end{figure*}
%%%%%%%%%%%%%%%%%%%%%%%%%%%%%%%%%%%%%%%%%%%%%%%%%%%%%%%%%%%%%%%%%%%%%%%%%%%
%%%%%%%%%%%%%%%%%%%%%%%%%%%%%%%%%%%%%%%%%%%%%%%%%%%%%%%%%%%%%%%%%%%%%%%%%%%
\begin{figure}
%\centering
\includegraphics[width=8cm, height=7cm]{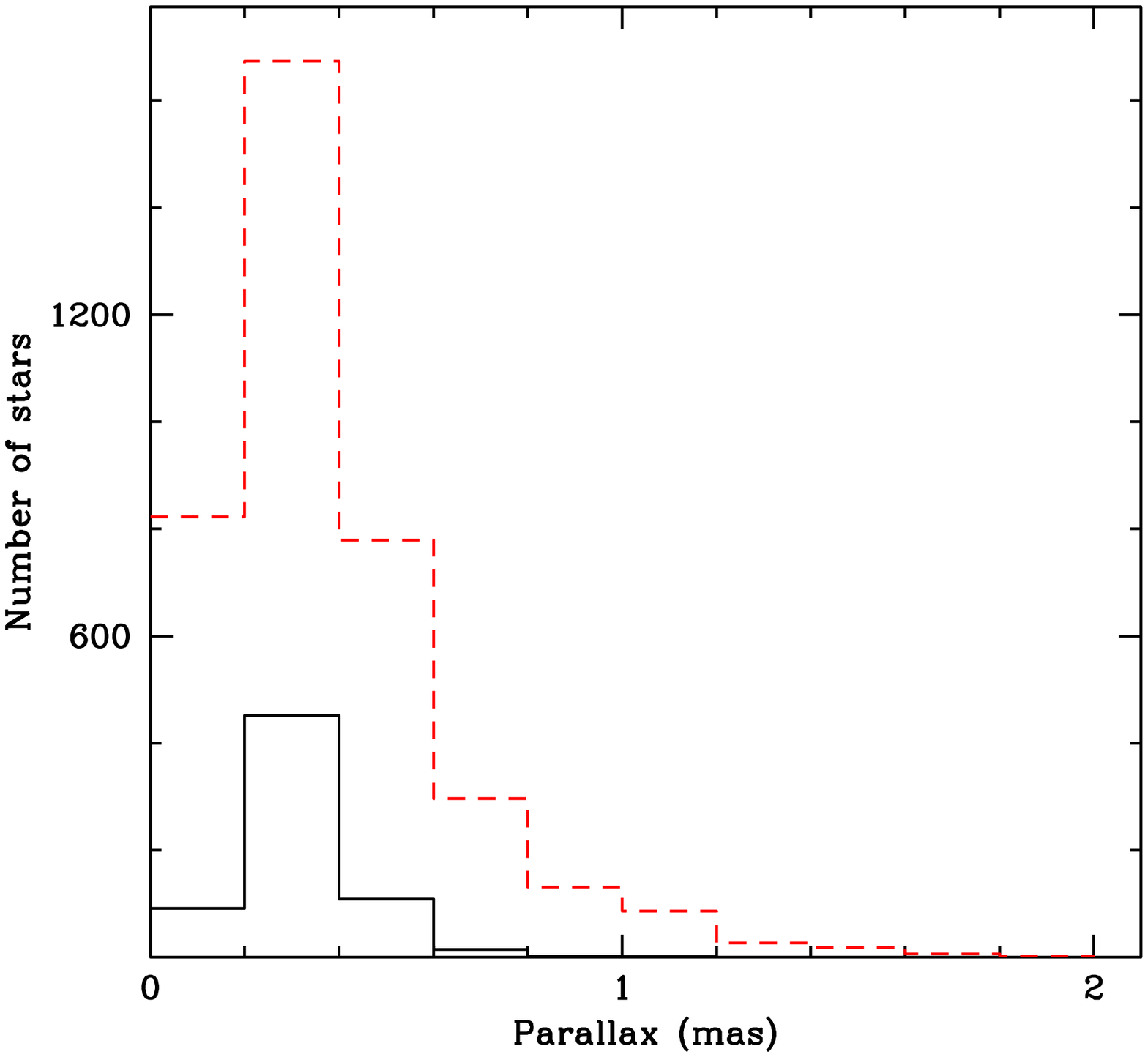}
\includegraphics[width=8cm, height=7cm]{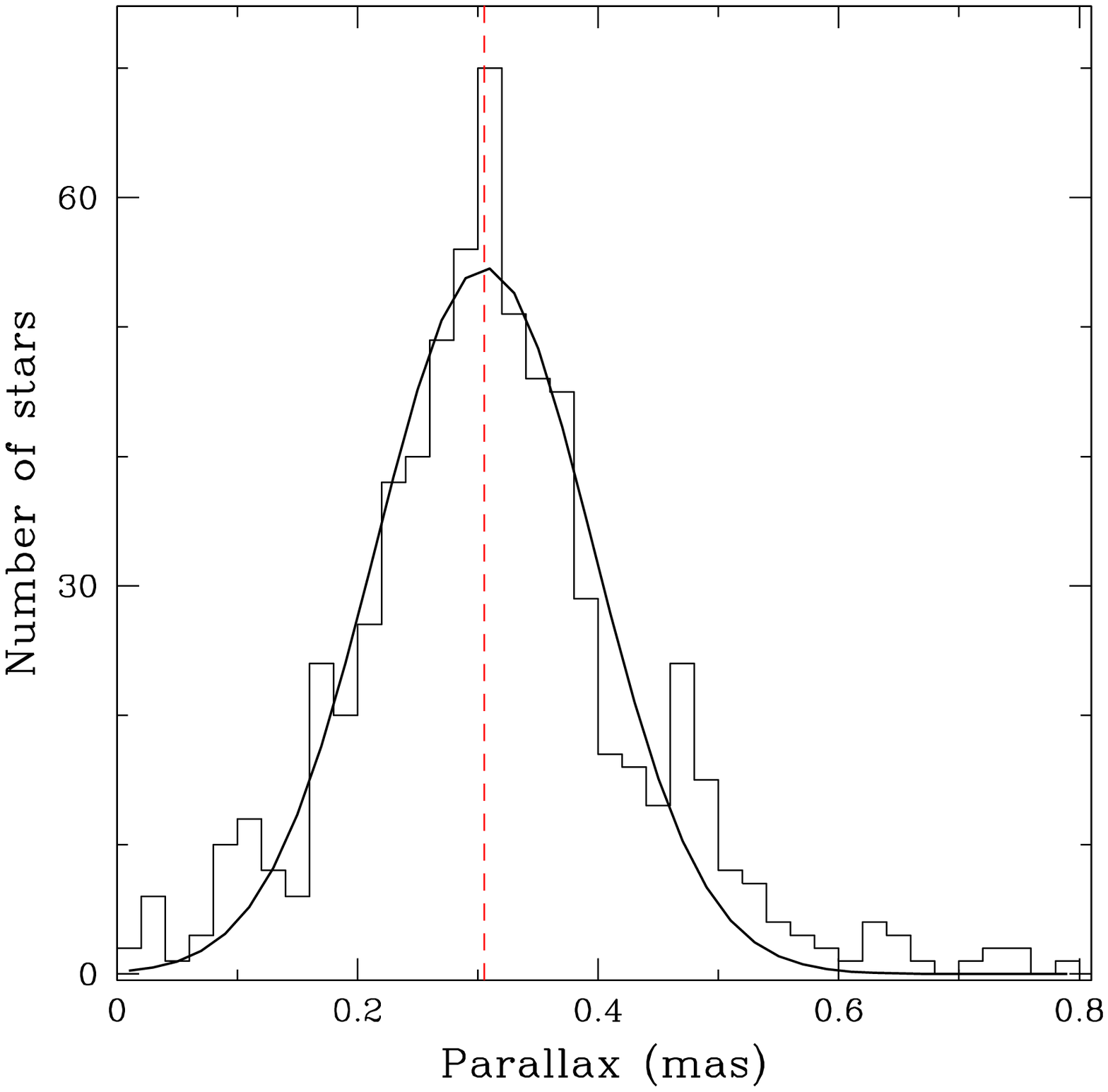}
\caption{
(Left) Shown here is the histogram of Gaia parallaxes in the cluster's region.
The dashed red line shows all the stars in our membership catalogue
while the most probable cluster members are shown as the solid black histogram. 
(Right) The mean parallax value calculated by fitting a Gaussian function. 
}
\label{figplx}
\end{figure}
%%%%%%%%%%%%%%%%%%%%%%%%%%%%%%%%%%%%%%%%%%%%%%%%%%%%%%%%%%%%%%%%%%%%%%%%%%%
%%%%%%%%%%%%%%%%%%%%%%%%%%%%%%%%%%%%%%%%%%%%%%%%%%%%%%%%%%%%%%%%%%%%%%%%%%%%%%%%%%%%%%%%%%%%%%%%
\begin{table*}
\centering
\caption{Shown here is the number of BSS in different radial bins for King 11. 
}
\begin{tabular}{cc}
\hline
\hline
Radial bin  & Number of BSS         \\
(arcmin)    &   \\
\hline
0--4     &   5  \\
4--8     &   2  \\
8--12    &   2  \\
12--16   &   3  \\
16--20   &   1  \\
\hline
\label{bss_radialdist}
\end{tabular}
\end{table*}
%%%%%%%%%%%%%%%%%%%%%%%%%%%%%%%%%%%%%%%%%%%%%%%%%%%%%%%%%%%%%%%%%%%%%%%%%%%%%%%%%%%%%%%%%%%%%%%%%%%%%%%%%%%%%%%%%%%%%%%%%%%%%%%%%%%%%%%%%%%%%%%%%%%%%%%%%%%%%%%%
\begin{figure*}
\begin{center}
\includegraphics[width=9.5cm, height=10.0cm]{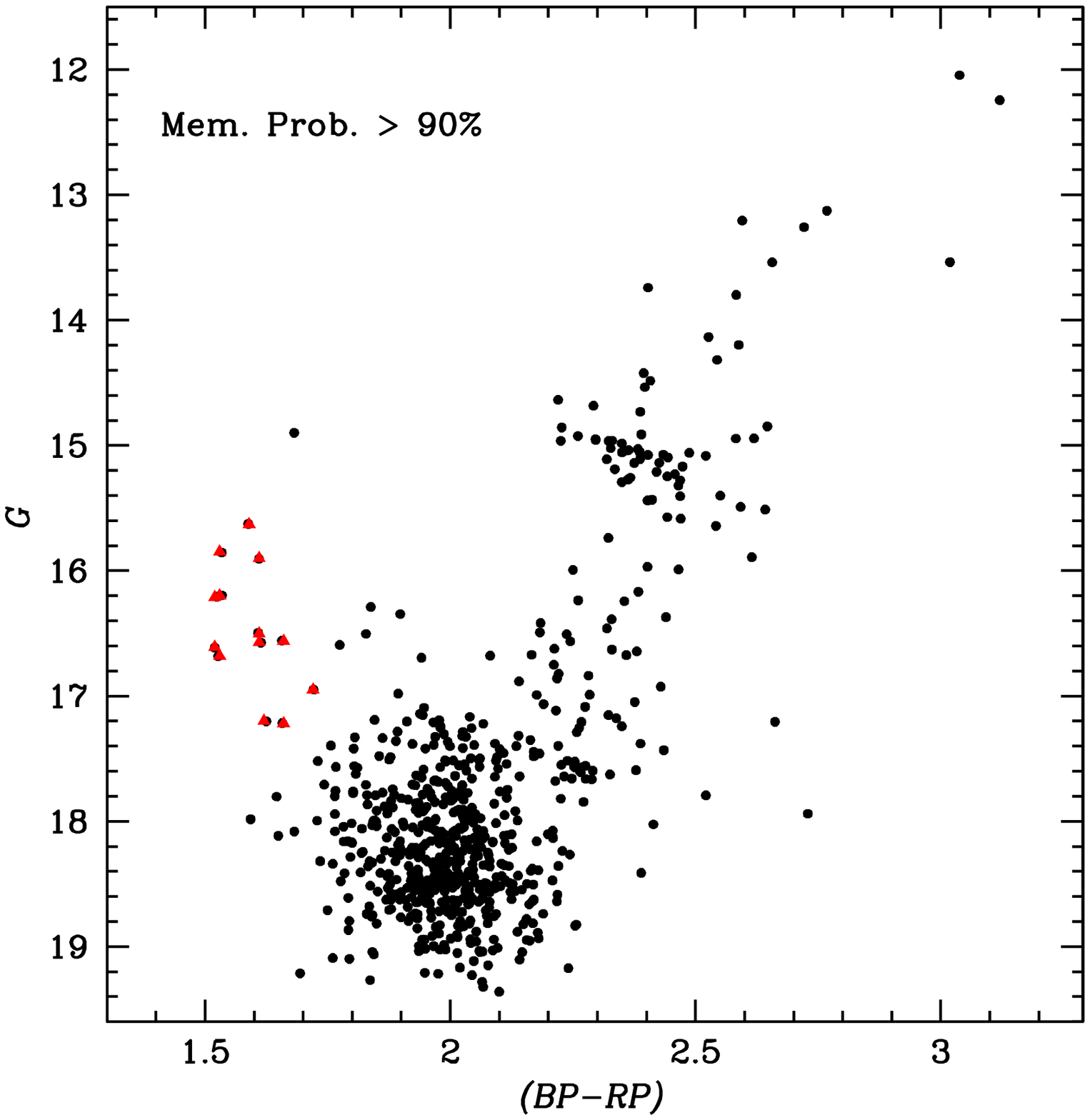}
\caption{
The location of the BSS population is shown here 
by the red triangles in the CMD of the most probable members of King 11.
}
\label{bssposition}
\end{center}
\end{figure*}
%%%%%%%%%%%%%%%%%%%%%%%%%%%%%%%%%%%%%%%%%%%%%%%%%%%%%%%%%%%%%%%%%%%%%%%%%%%

\subsection{Membership probabilities}
\label{MP}

With the availability of the precise PMs from Gaia-EDR3, we aim to do more than 
just the preliminary separation of field stars which was shown in Section~\ref{VPD}.
The mathematical set up for the membership probability for the star clusters
was presented by Vasilevskis et al. (1958). 
In this paper, we use the method devised by Balaguer-N\'{u}\~{n}ez et al. (1998)
to calculate the membership probability of individual stars in the area of King 11.
This method has been used previously by our group for both open and globular
star clusters (Yadav et al. 2013, Sariya \& Yadav 2015; Sariya et al. 2021a, 2021b) 
and a detailed description of this method can be found in Bisht et al (2020).

To derive the two distribution functions defined in this method,
$\phi_c^{\nu}$ (cluster star distribution) 
and $\phi_f^{\nu}$ (field star distribution), 
we considered only those stars which have PM errors better than 1 mas~yr$^{-1}$
and parallax errors $<$1 mas. 
A group of the cluster's preliminary members shown in the VPD
is found to be centered at 
$\mu_{xc}$=$-$3.39 mas~yr$^{-1}$, $\mu_{yc}$=$-$0.675 mas~yr$^{-1}$ .
We have estimated the PM dispersion for the cluster population as 
($\sigma_c$) = 0.1 $mas~yr^{-1}$. 
For the field region, we have estimated 
($\mu_{xf}$, $\mu_{yf}$) = ($-$1.94, $-$0.39) mas yr$^{-1}$ 
and ($\sigma_{xf}$, $\sigma_{yf}$) = (0.98, 0.68) mas yr$^{-1}$.\\

The membership probabilities ($P_{\mu}$) are thus determined and they are shown 
as a function of the Gaia's $G$ magnitude 
in the left panel of Fig.~\ref{mp_hist}.
We plotted a histogram of membership probabilities in the right panel of 
this figure. Based on the histogram, we adopted the stars with 
$P_{\mu}>$ 90\% as the most probable cluster members. 
This cut-off of 90\% is also shown as a horizontal dashed line in the 
left panel of Fig.~\ref{mp_hist}.
Finally, we identified a total of 676 stars with $P_{\mu}>$ 90\%
which also lie within the cluster's limiting radius (see Section~\ref{RADIUS}).
These stars are plotted in the left panel of Fig.~\ref{figmp90}.

Cantat-Gaudin et al. (2018) had also determined the membership of stars in King 11.
The stars with $P_{\mu}>$ 90\% in the catalogue from Cantat-Gaudin et al. (2018)
are shown in a CMD plotted in the right panel of Fig.~\ref{figmp90}.
Their catalogue goes only up to $\sim$18 $G$ mag and covers about 10 arcmin radius
of the cluster.
Our membership catalogue covers a radius of $\sim$20 arcmin and reaches deeper
in magnitude ($G \sim$ 19.6 mag).

To estimate the mean PM of King 11, 
we considered the most probable cluster members
and constructed the histograms for 
$\mu_{\alpha} cos{\delta}$ and $\mu_{\delta}$ 
as shown in Fig.~\ref{pm_hist}. 
The fitting of a Gaussian function to the histograms provides 
the mean PM in both directions. 
We obtained the mean PM of King 11 as 
$-3.391\pm0.006$ and $-0.660\pm0.004$ mas yr$^{-1}$ in
$\mu_{\alpha} cos{\delta}$ and $\mu_{\delta}$ respectively. 
The estimated values of mean PM for this object 
is in very good agreement with the mean PM value given by 
Cantat-Gaudin et al. (2018) which is mentioned in Section~\ref{INTRO}.

The left panel of Fig.~\ref{figplx} shows the histograms of 
all the stars and the most probable cluster members. 
In the right panel of this figure, a Gaussian fit to the 
histograms of parallax of the most probable cluster members 
provide a mean parallax value of 
0.306 $\pm$ 0.004 mas. 

%%%%%%%%%%%%%%%%%%%%%%%%%%%%%%%%%%%%%%%%%%%%%%%%%%%%%
\subsubsection{The BSS of King 11}

The BSS are important to understand the theory of stellar evolution.
Sandage (1953) was the first to point out their presence in a globular cluster.
Two main mechanisms are proposed for their formation:
(i) the mass-transfer in a binary stellar system 
(McCrea 1964; Zinn \& Searle 1976), and
(ii) stellar merger owing to the collisions (Hills \& Day 1976).

Based on the radial distribution of the BSS in globular clusters,
Ferraro et al. (2012) defined three classes:
family I for a flat distribution, 
family II for a bimodal radial distribution, and
family III if the distribution of the BSS  
peaks in the central region of the cluster.
Now, there are papers where this analogy is being applied 
to the BSS in open clusters as well (e.g. Vaidya et al. 2020).
For Berkeley 17, the BSS distribution was found to be 
similar to the family II globular clusters (Bhattacharya et al. 2019).
Rain et al. (2020) found a flat distribution (family I)
for Collinder 261.

We used the visual inspection method of the CMD 
along with the criterion given by McCrea (1964)
to pick out the cluster's BSS population. 
We detected 13 BSS as the most probable members of the 
open cluster King 11.
In Fig.~\ref{bssposition}, the BSS of King 11 
are shown in the cluster's CMD by the notation of red triangles.
The radial distribution of the BSS is presented in Table~\ref{bss_radialdist}.
It is obvious from the Table that the maximum number of BSS 
are located in the central bin, while the distribution is 
roughly uniform after that.
Based on the above discussion,
King 11 can be considered similar to the family III
clusters. 

As shown in Fig.~\ref{rdp}, the cluster has a 
higher number density of stars in the central area.
Thus, the presence of the maximum number of the BSS 
in the central area can be connected with a higher density of stars 
in that region.
The higher density enhances the possibility of occurrence 
for the mechanisms responsible for the formation of the BSS.

\section{Structural and fundamental parameters of King 11}
\label{FUNDA}

We discuss the determination of several parameters in this section. 
The available literature values of various parameters for King 11
are compared with the present analysis in Table~\ref{complit}.

%%%%%%%%%%%%%%%%%%%%%%%%%%%%%%%%%%%%%%%%%%%%%%%%%%%%%%%%%%%%%%%%%%%%%%%%%%
\begin{table*}
{\footnotesize
\tabcolsep=0.15cm
\caption{
The results of present work compared with the published literature.
}
\vspace{0.5cm}
\centering
\begin{tabular}{lcl}
\hline\hline
\noalign{\smallskip}
Parameters & Value & Reference \\
\hline
\noalign{\smallskip}
Age, Gyr             &$ 3.63\pm0.42$           &Present work \\
                     & 2.09                    &Kharchenko et al. (2013)\\
                     & 3.00$\pm$ 0.36           &Kyeong et al. (2011)\\
                     &$3.4$ -- $ 4.75$         &Tosi et al. (2007)\\
                     &$ \sim5\pm1$             &Aparicio et al. (1991) \\
                     &$ \sim5$                 &Kaluzny (1989)\\
Z$_{metal}$          & 0.011                   &Present work\\
                     &$ 0.01$                  &Tosi et al. (2007)\\
                     &$ 0.02$                  &Aparicio et al. (1991)\\
distance, kpc        &$ 3.33\pm0.15$           &Present work\\
                     & 3.433                   &Cantat-Gaudin et al. (2018)\\
                     & 2.850                   &Kharchenko et al. (2013)\\
                     & 2.19                    &Friel et al. (2002)\\
($\mu_\alpha\cos\delta, \mu_\delta$), mas yr$^{-1}$ & $(-3.391\pm0.006$, $ -0.660\pm0.004$) & Present work  \\
                     & ($-$3.34, $-$ 0.60)   & Liu \& Pang (2019)  \\
                     & ($-$3.358, $-$ 0.643)   &Cantat-Gaudin et al. (2018)  \\
$V_r$, km s$^{-1}$   &$ -24.96\pm5.00$         &Present work\\
                     &$ -24.61\pm0.34$         &Soubiran et al. (2018)\\
                     &$ -35$                   &Kharchenko et al. (2013)\\
                     &$ -35\pm16$              &Friel et al. (2002)\\
                     &$ -34\pm12$              &Scott et al. (1995)\\
plx, mas             & 0.306$\pm$ 0.004         &Present work   \\
                     & 0.299                   & Liu \& Pang (2019) \\
                     & 0.262                   &Cantat-Gaudin et al. (2018) \\
$Z_{0}$, kpc         &$ 0.401\pm0.010$         &Present work\\
                     & 0.4017                  &Soubiran et al. (2018)\\
                     & 0.3877                  &Cantat-Gaudin et al. (2018)\\
                     & 0.253 -- 0.387          &Tosi et al. (2007)\\
                     & 0.245                   &Friel et al. (2002)\\
\hline
\label{complit}
\end{tabular}
}
\end{table*}
%%%%%%%%%%%%%%%%%%%%%%%%%%%%%%%%%%%%%%%%%%%%%%%%%%%%%%%%%%%%%%%%%%%%%%%%%%%

\subsection{Radial density profile}
\label{RADIUS}

\begin{figure*}
\begin{center}
\includegraphics[width=7.5cm, height=7.5cm]{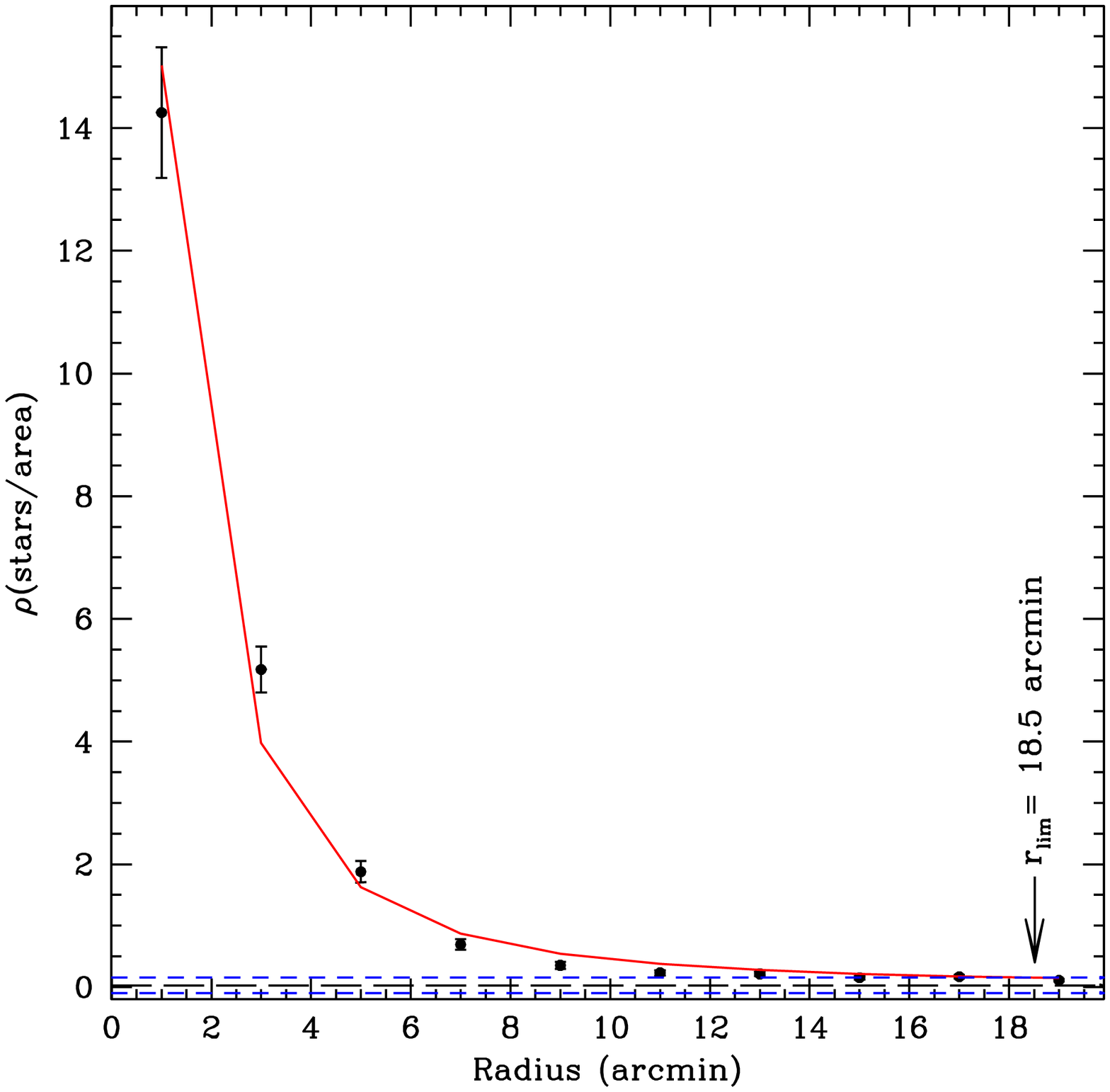}
\caption{The stellar number density distribution of the most probable
cluster members for the open cluster King 11. 
The fitted curve represents the King (1962) profile.
The horizontal dashed lines show the level of $f_{bg}$
with 3$\sigma$ errors.
}
\label{rdp}
\end{center}
\end{figure*}

To know about the extent of the cluster, 
we plotted the radial density profile (RDP) as shown in Fig.~\ref{rdp}.
We divided the area of King 11 into many concentric rings. 
The stellar number density, $R_{i}$, in the $i^{th}$ zone 
is determined by using the formula:
$R_{i}$ = $\frac{N_{i}}{A_{i}}$,
where $N_{i}$ is the number of stars 
and $A_{i}$ is the area of the $i^{th}$ zone. 
A smooth continuous line represents the fitted King (1962) profile:\\

\begin{equation}
f(r) = f_{bg}+\frac{f_{0}}{1+(r/r_{c})^2}\\
\end{equation} 

where $r_{c}$, $f_{0}$, and $f_{bg}$ are the 
core radius, central density, and the background density level, respectively.
By fitting the King model to the radial density profile, 
we estimated the structural parameters for King 11. 
The obtained values of
$r_{c}$, $f_{0}$ and $f_{bg}$ are:
$1.36\pm0.24$ arcmin,
$23.02\pm5.86$ stars per arcmin$^{2}$
and $0.03\pm0.04$ stars per arcmin$^{2}$.
The levels of $f_{bg}$ and its 3$\sigma$ errors 
are also shown by the dashed lines in Fig.~\ref{rdp}. 
To calculate the limiting radius ($r_{lim}$) of the cluster, we used 
the relation mentioned by Bukowiecki et al. (2011) as:
$r_{lim}=r_{c}\sqrt(\frac{f_{0}}{3\sigma_{bg}}-1)$.
Thus, for King 11, we obtained a value of $r_{lim}$ = 18.51$^{\prime}$.

%%%%%%%%%%%%%%%%
%%%%%%%%%%%%%%%%

\subsection{Age and distance}
\label{ISOCHRONE}

%%%%%%%%%%%%%%%%
\begin{figure*}
\begin{center}
\includegraphics[width=8.5cm, height=8.5cm]{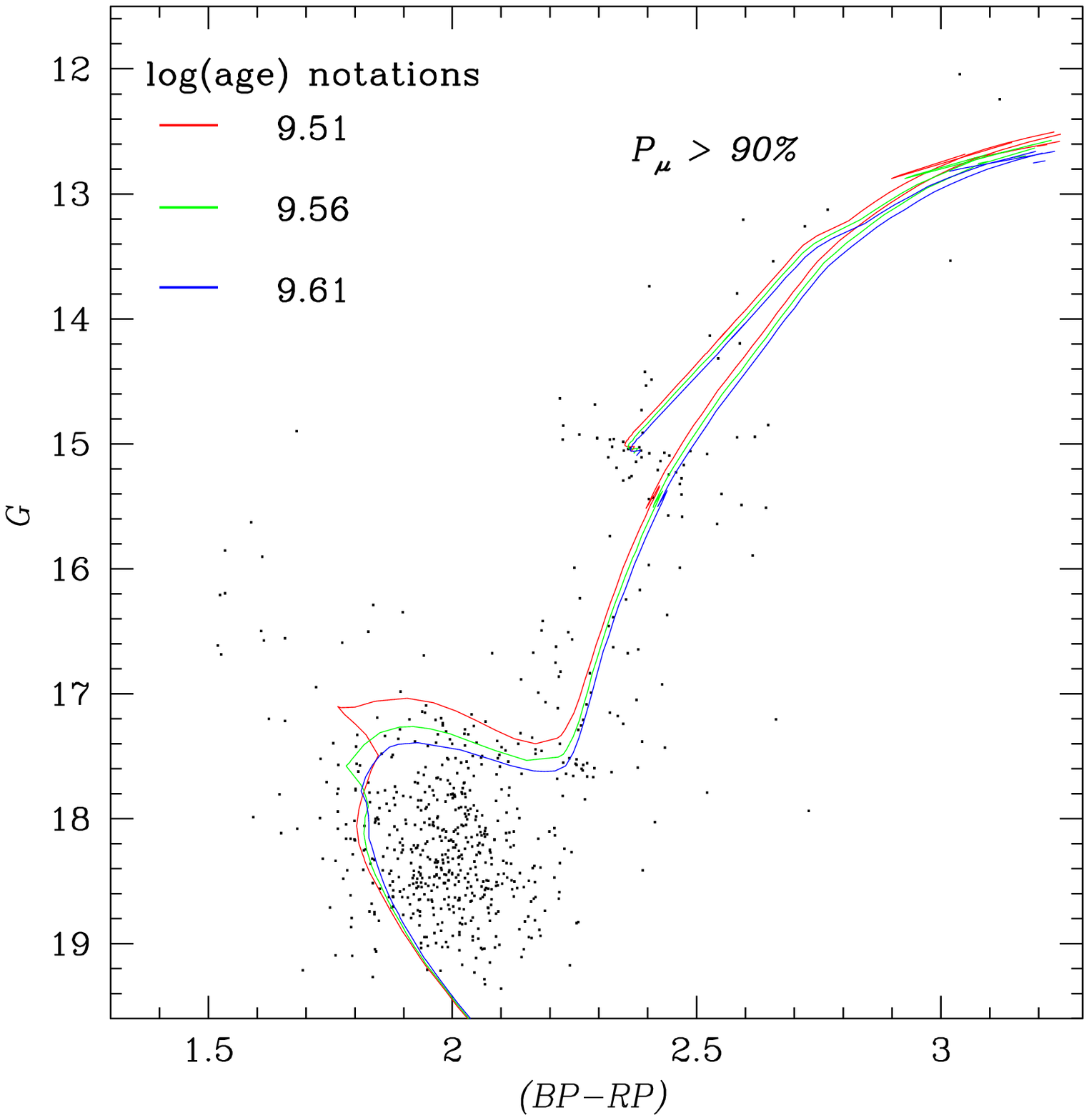}
\caption{Isochrones given by Bressan et al. (2012) fitted to 
the $G, (BP-RP)$ CMD of the most probable members of King 11. 
These isochrones have Z$_{metal}$ = 0.011 and log(age) values 
of 9.51, 9.56 and 9.61.
}
\label{age}
\end{center}
\end{figure*}
%%%%%%%%%%%%%%%%

The ages and distances of the old open star clusters can be used to 
trace the structure and chemical evolution of the Galaxy (Friel \& Janes 1993). 
The astrophysical parameters like the metallicity, age, distance and reddening
of King 11 are estimated by fitting 
the theoretical evolutionary isochrones of Bressan et al. (2012) 
to the observed CMDs as shown in Fig.~\ref{age}. 
The superimposed isochrones are of different age values 
(log(age)=9.51, 9.56 and 9.61) and have a metallicity of Z$_{metal}$=0.011. 
Thus, the age of King 11 is obtained as $3.63\pm0.42$ Gyr. 
The present values of metallicity and age 
are quite similar to the values (Z$_{metal}$=0.01, age=3.5-4.75 Gyr)
suggested by Tosi et al. (2007). 
The estimated distance modulus is ($m-M_{G}$)= 14.65$\pm$0.02 mag.
We used the extinction relations given by Hendy (2018) to convert
the distance modulus to the heliocentric distance of the cluster. 
Thus, the present study presents 
a heliocentric distance of King 11 as $3.33\pm0.15$ kpc.
This value is in agreement with the distance (3.43 kpc) given by 
Cantat-Gaudin et al. (2018).
The visual fitting also provided a high reddening 
($E(BP-RP)$ = 1.19) for the cluster.

%%%%%%%%%%%%%%%%%%%%%%%%%%%%%%%%%%%%%%%%%%%%%%%%%%%%%%%%%%%%%%%%%%%%%%%%%%%
\begin{figure}
\centering
\includegraphics[width=8.5cm, height=8.5cm]{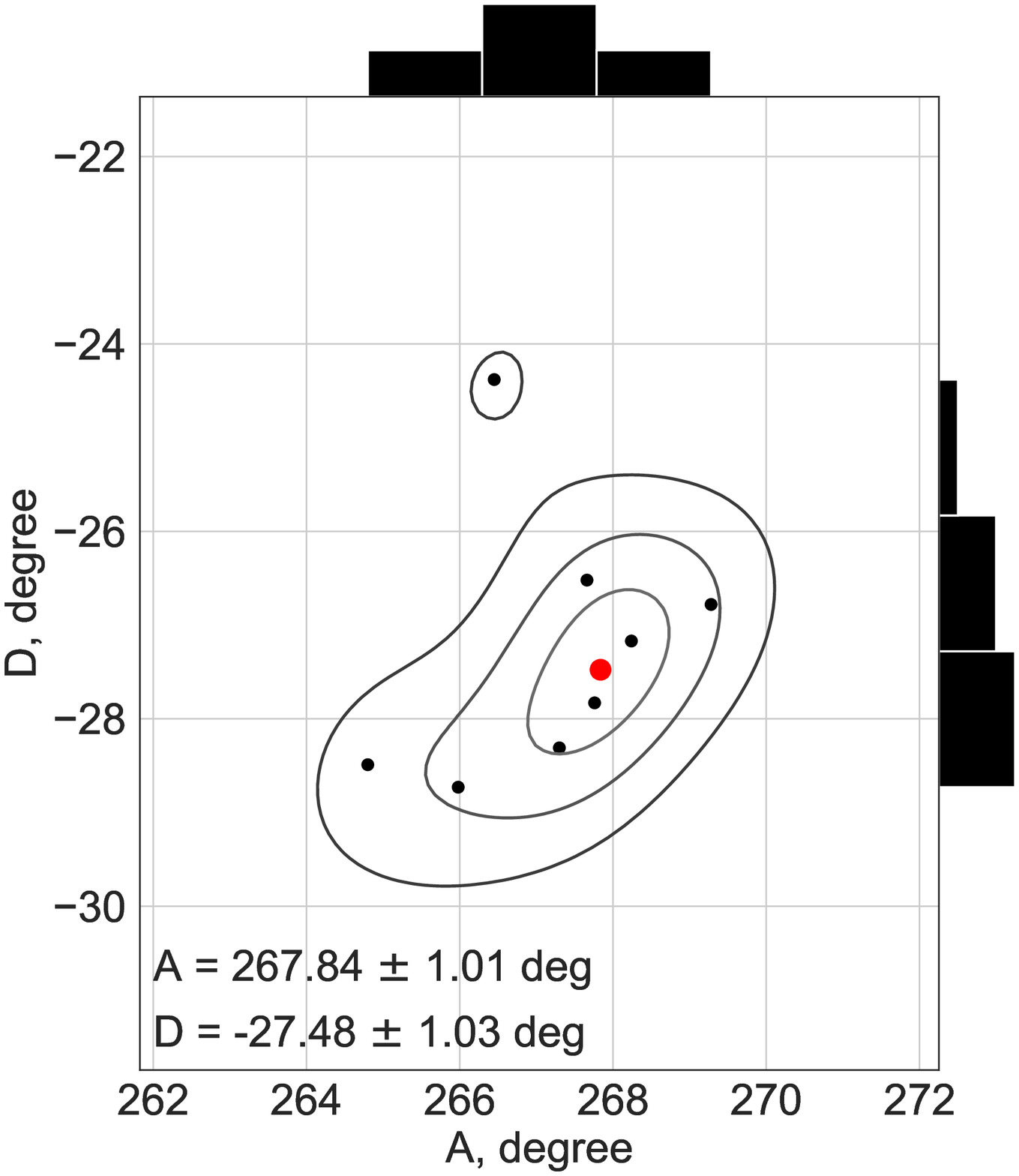}
\caption{
The AD plotted for King~11. We used n=8 stars for which 
we managed to find all the parameters necessary for calculating the apex.
The red dot in the central contour indicates the mean apex for the cluster. 
}
\label{fig_AD}
\end{figure}
%%%%%%%%%%%%%%%%%%%%%%%%%%%%%%%%%%%%%%%%%%%%%%%%%%%%%%%%%%%%%%%%%%%%%%%%%%%

\section{The apex of King 11}
\label{adapex}

%%%%%%%%%%%%%%%%%%%%%%%%%%%%%%%%%%%%%%%%%%%%%%%%%%%%%%%%%%%%%%%%%%%%%%%%%%
\begin{table*}
{\footnotesize
\tabcolsep=0.1cm
\caption{
Data for n=9 most probable cluster member stars with the available radial velocity values in Gaia-EDR3.
}
\vspace{0.5cm}
\centering
\begin{tabular}{ccccccccccc}
\hline\hline
\noalign{\smallskip}
	Our ID & $\alpha$ (J2000)& $\delta$ (J2000) &$\pi$ &$\sigma_\pi$ &$V_r$ &$\sigma_{V_r}$ &
$\mu_\alpha cos{\delta}$&$\sigma_{\mu_\alpha cos{\delta}}$&$\mu_\delta$&$\sigma_{\mu_\delta}$\\
	& deg      & deg       & mas  & mas         & km s$^{-1}$ &km s$^{-1}$     &
 mas yr$^{-1}$      & mas yr$^{-1}$        & mas yr$^{-1}$   & mas yr$^{-1}$               \\
\hline
\noalign{\smallskip}
59&356.94999&68.648796&0.304&0.013& $-$27.33 &0.73&$-$3.373&0.015&$-$0.549&0.014\\
71&356.99190&68.627326&0.302&0.021& $-$27.65 &11.8&$-$3.408&0.024&$-$0.625&0.025\\
186&356.91987&68.608428&0.302&0.011& $-$23.00 &1.14&$-$3.458&0.012&$-$0.545&0.013\\
188&356.91083&68.656766&0.307&0.020& $-$23.75 &0.4&$-$3.377&0.022&$-$0.751&0.022\\
383&357.07390&68.619668&0.315&0.010& $-$24.36 &0.44&$-$3.329&0.012&$-$0.678&0.012\\
402&357.08317&68.624852&0.315&0.012& $-$23.37 &1.24&$-$3.303&0.013&$-$0.680&0.014\\
594&356.76670&68.616595&0.303&0.011& $-$25.56 &0.61&$-$3.287&0.012&$-$0.674&0.013\\
664&356.89949&68.559192&0.313&0.017& $-$24.64 &0.19&$-$3.548&0.018&$-$0.691&0.020\\
750&356.81563&68.703584&0.264&0.013& $-$114.10&0.53&$-$3.464&0.014&$-$0.612&0.014\\
\hline
\label{tab_forAD}
\end{tabular}
}
\end{table*}
%%%%%%%%%%%%%%%%%%%%%%%%%%%%%%%%%%%%%%%%%%%%%%%%%%%%%%%%%%%%%%%%%%%%%%%%%%%

%%%%%%%%%%%%%%%%%%%%%%%%%%%%%%%%%%%%%%%%%%%%%%%%%%%%%%%%%%%%%%%%%%%%%%%%%%%
\begin{table*}
{\footnotesize
\tabcolsep=0.1cm
\caption{
The obtained values of the positions of the individual apexes for the stars of King 11. 
}
\vspace{0.5cm}
\centering
\begin{tabular}{cccccccc}
\hline\hline
\noalign{\smallskip}
Our ID&B-J distance & B-J lower &B-J upper &  $d$=$1/\pi$ &relative parallax &$A$ & $D$ \\
	& kpc       & kpc & kpc & kpc & error (\%)    & deg & deg\\
\hline
\noalign{\smallskip}
 59&3.2259&2.8823&3.6581&3.289&4.4&264.80&$-$28.49\\
 71&3.8800&3.3699&4.5585&3.311&7.1&265.98&$-$28.73\\
186&3.9976&3.5189&4.6171&3.301&3.6&266.45&$-$24.38\\
188&4.8939&3.9317&6.3448&3.257&6.6&269.28&$-$26.78\\
383&3.1930&2.8330&3.6526&3.169&3.3&267.76&$-$27.83\\
402&3.0465&2.6475&3.5786&3.172&4.0&268.24&$-$27.17\\
594&4.9844&3.9980&6.4726&3.292&3.9&267.30&$-$28.31\\
664&4.2290&3.6042&5.0915&3.187&5.4&267.66&$-$26.52\\
\hline                                    
\label{tab_AD}
\end{tabular}
}
\end{table*}
%%%%%%%%%%%%%%%%%%%%%%%%%%%%%%%%%%%%%%%%%%%%%%%%%%%%%%%%%%%%%%%%%%%%%%%%%%%

To study the stellar population of  Hyades, van Altena (1969) developed
the classic method of analyzing PMs of the cluster members.
In contrast, the apex diagram (AD) method includes the radial velocity
along with the PMs. Clearly, it affects the number of stars 
available for the analysis
due to the limited availability of the stars with radial velocity measurements.
However, the AD-method allows us to study the kinematical stellar structures within the cluster
and presents a more complete picture of the motion of stars
using the spatial velocities of stars.
The details of the AD-method can be found in Chupina et al. (2006).
We used this method to study the kinematics of various open clusters, 
namely M~67 (Vereshchagin et al. 2014), Pleiades (Elsanhoury et al. 2018), 
IC~2391 (Postnikova et al. 2020) and NGC~2158 (Sariya et al. 2021a).

The coordinates $(A,D)$ for a star represent a point on the celestial sphere
in a rectangular heliocentric coordinate system where $A$ is for the right ascension 
and $D$ represents the declination.
The $(A,D)$ set the direction of the spatial velocity vectors.
The close locations of the $(A,D)$ for the cluster members in AD-chart 
signifies the similar directions of the corresponding vectors in space.
The commonality of motion distinguishes the cluster member stars 
from the background stars.
The individual stellar apexes are scattered around the mean cluster apex.
Thus, the cluster's apex can be calculated by taking mean of the 
individual apexes of the most probable cluster members. 

Among the 676 most probable cluster members, only 9 stars
have the availability of radial velocity values in Gaia database. 
We did not find radial velocities in the 
Large Sky Area Multi-Object Fibre Spectroscopic Telescope 
(LAMOST) DR5 (Luo et al. 2019) 
for this cluster because the declination of King 11 
exceeds the area covered by the LAMOST.

The data for these 9 stars is provided in Table~\ref{tab_forAD}.
The columns of Table~\ref{tab_forAD} contain:
the star ID number in our membership catalogue, 
the equatorial coordinates ($\alpha, \delta$) in J2000, 
parallaxes and their errors ($\pi$, $\sigma_\pi$), 
radial velocities and their errors ($V_r, \sigma_{V_r}$), km s$^{-1}$,
PMs with their errors along RA 
($\mu_\alpha cos{\delta}, \sigma_{\mu_\alpha cos{\delta}}$),
and DEC ($\mu_\delta, \sigma_{\mu_\delta}$) in mas yr$^{-1}$.

To calculate the position of the apexes of the stars, 
we used the data from Table~\ref{tab_forAD}.
Star with ID=750 was excluded from the apex calculations because
its radial velocity value falls beyond 3$\sigma$ 
of the cluster's average radial velocity.
Thus, we used n=8 stars for apex determination. 
The apex calculation requires the distance of the stars from the Sun.
In Table~\ref{tab_forAD}, we have the Gaia-EDR3 parallax values of the stars.
We had two options, whether to use the distances provided by 
Bailer-Jones et al. (2018, B-J distances) based on Gaia DR2
or to use the reciprocal of the Gaia-EDR3 parallaxes which are 
more precise than Gaia-DR2 parallaxes.
The possibility of using the parallax reciprocal depends 
on the value of the relative parallax error (Luri et al. 2018).
For instance, for another distant cluster $h$ and $\chi$ Persei,
the parallax inversion was done 
when the relative parallax error was less than 15\% (Zhong et al. 2019).
In the present work, as can be seen in Table~\ref{tab_AD}, 
the relative parallax errors do not exceed $7$\%.
Also, it is evident from Table~\ref{tab_AD}, 
that as compared to the B-J distances,
the distances based on $1/\pi$ (where the parallaxes are from Gaia-EDR3) 
formula provides distances with a smaller dispersion.
In addition, these distances are much closer to the 
distance value obtained by isochrone fitting in Section~\ref{ISOCHRONE}. 
Hence, we used the inverse of Gaia-EDR3 parallaxes for 
calculating the individual stellar distances.  
Table~\ref{tab_AD} shows the resulting apex coordinates.
The columns of Table~\ref{tab_AD} contain:
the star ID number, B-J distance,
the lower and upper limits of the B-J distance,
the distance calculated by $1/\pi$,
the relative parallax error (\%), 
and the coordinates of the individual apex positions.
The mean apex was determined as:
($A, D$) = $267.84^\circ\pm1.01^\circ$, $-27.48^\circ\pm1.03^\circ$.
Figure~\ref{fig_AD} shows the positions of the apexes of the stars
in the equatorial coordinate system,
as well as the position of the mean apex for the cluster

\subsection{Diagram of $\mu_U$ and $\mu_T$}
\label{secmuut}

%%%%%%%%%%%%%%%%%%%%%%%%%%%%%%%%%%%%%%%%%%%%%%%%%%%%%%%%%%%%%%%%%%%%%%%%%%%
\begin{figure}
%\centering
\includegraphics[width=\textwidth]{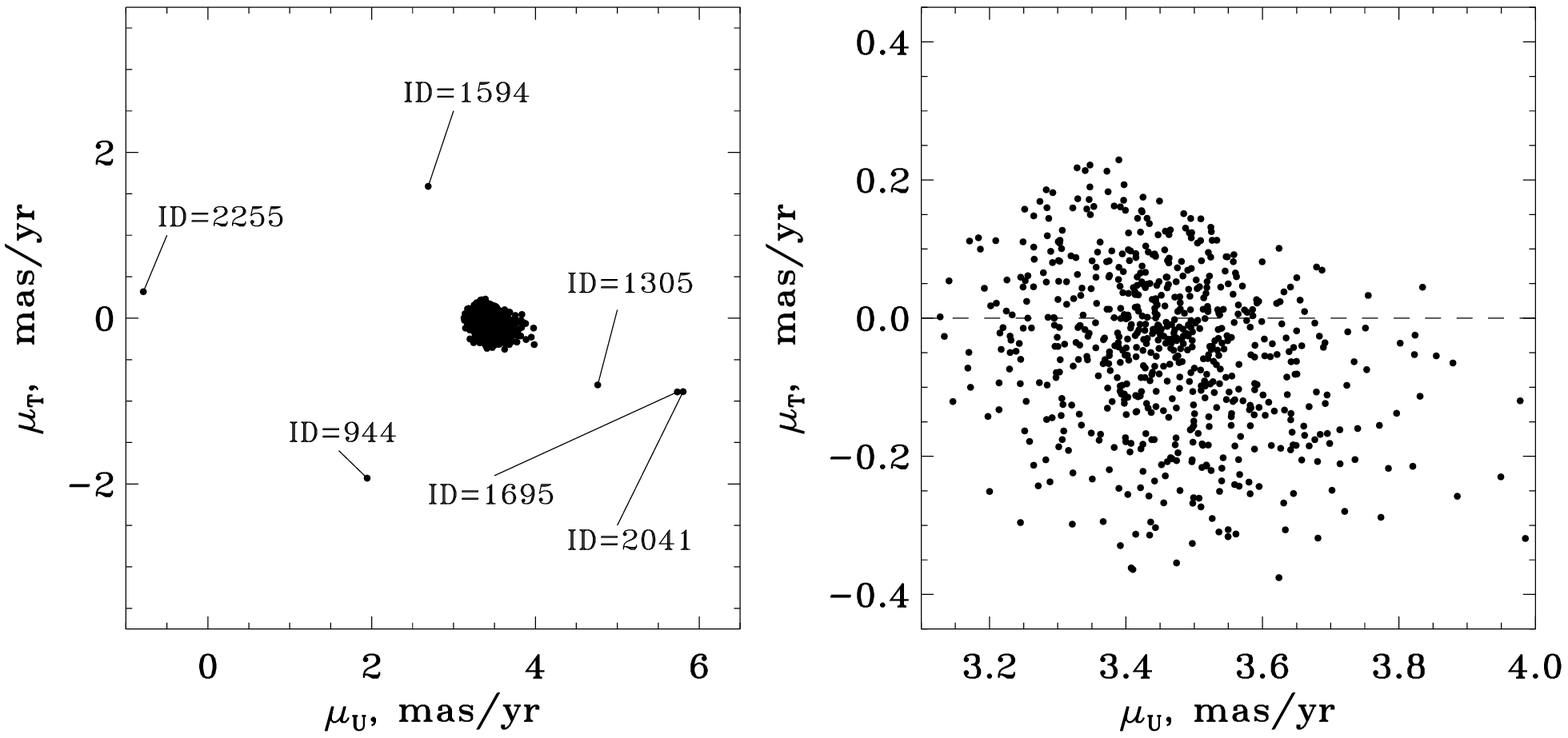}
\caption{
The ($\mu_U$, $\mu_T$) diagram for 676 most probable member stars
with $P_{\mu}>90\%$ and radial distance $<18.51$~arcmin.
In the left panel, we can see some stars that fall 
beyond the 3$\sigma$ from the mean value. 
In the right panel, a zoomed in version of the central concentration is shown.
}
\label{fig_muu_mut}
\end{figure}
%%%%%%%%%%%%%%%%%%%%%%%%%%%%%%%%%%%%%%%%%%%%%%%%%%%%%%%%%%%%%%%%%%%%%%%%%%%

As mentioned earlier, the PMs are available in a much larger number 
than the radial velocities. 
For King 11, we have 676 most probable cluster members with PM data.
These can be used to construct the ($\mu_U$, $\mu_T$) diagram. 
In principle, a preliminary assessment of the cluster's apex 
is also required here.
A special coordinate system is defined to construct this diagram
such that the reference axis 
$\mu_U$ is directed to a point on the celestial sphere 
representing the apex position, 
and the $\mu_T$ axis is directed perpendicular to the $\mu_U$ axis.
Thus, by the ($\mu_U, \mu_T$) diagram, 
the quality of the apex will be verified if we find that $\mu_T\approx0$.

For the most probable members of King 11, the resulting
($\mu_U, \mu_T$) diagram is presented in Figure~\ref{fig_muu_mut}.
In general, the distribution of stars in the figure is uniform.
The left panel of the figure shows some stars that are deviated from 
a set pattern by more than three sigma of the mean values. 
These 6 stars with their IDs mentioned in the figure were discarded.
A centralized distribution of the usable stars is shown 
on an enlarged scale in the right panel.
In order to numerically estimate the reliability of the 
obtained values of the cluster's apex in Section~\ref{adapex},
we calculated the average value of $\mu_T$ for 670 stars,
which is $\left<\mu_T\right>=-0.050\pm0.160$~mas yr$^{-1}$,
while $\left<\mu_U\right>=3.460\pm0.260$~mas yr$^{-1}$.
The value of mean $\mu_T$ being close to zero indicates
that the mean apex for King 11 determined in Section~\ref{adapex} is well defined.

\section{The cluster's orbit in the Galaxy}
\label{orbit}

%%%%%%%%%%%%%%%%%%%%%%%%%%%%%%%%%%%%%%%%%%%%%%%%%%%%%%%%%%%%%%%%%%%%%%%%%%%
\begin{figure}
%\centering
\includegraphics[height=8.5cm, width=10cm]{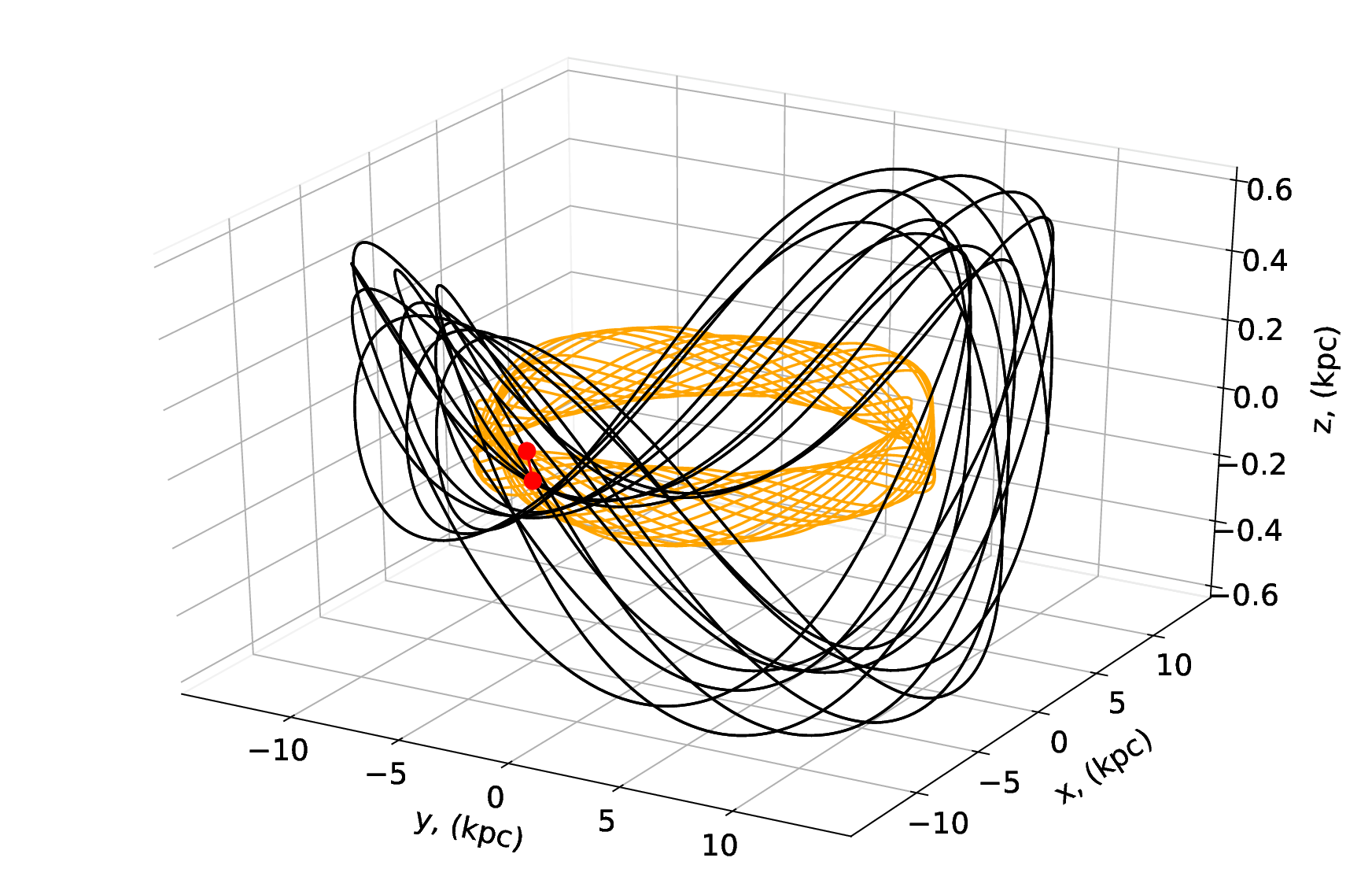}
\includegraphics[height=7.5cm, width=8cm]{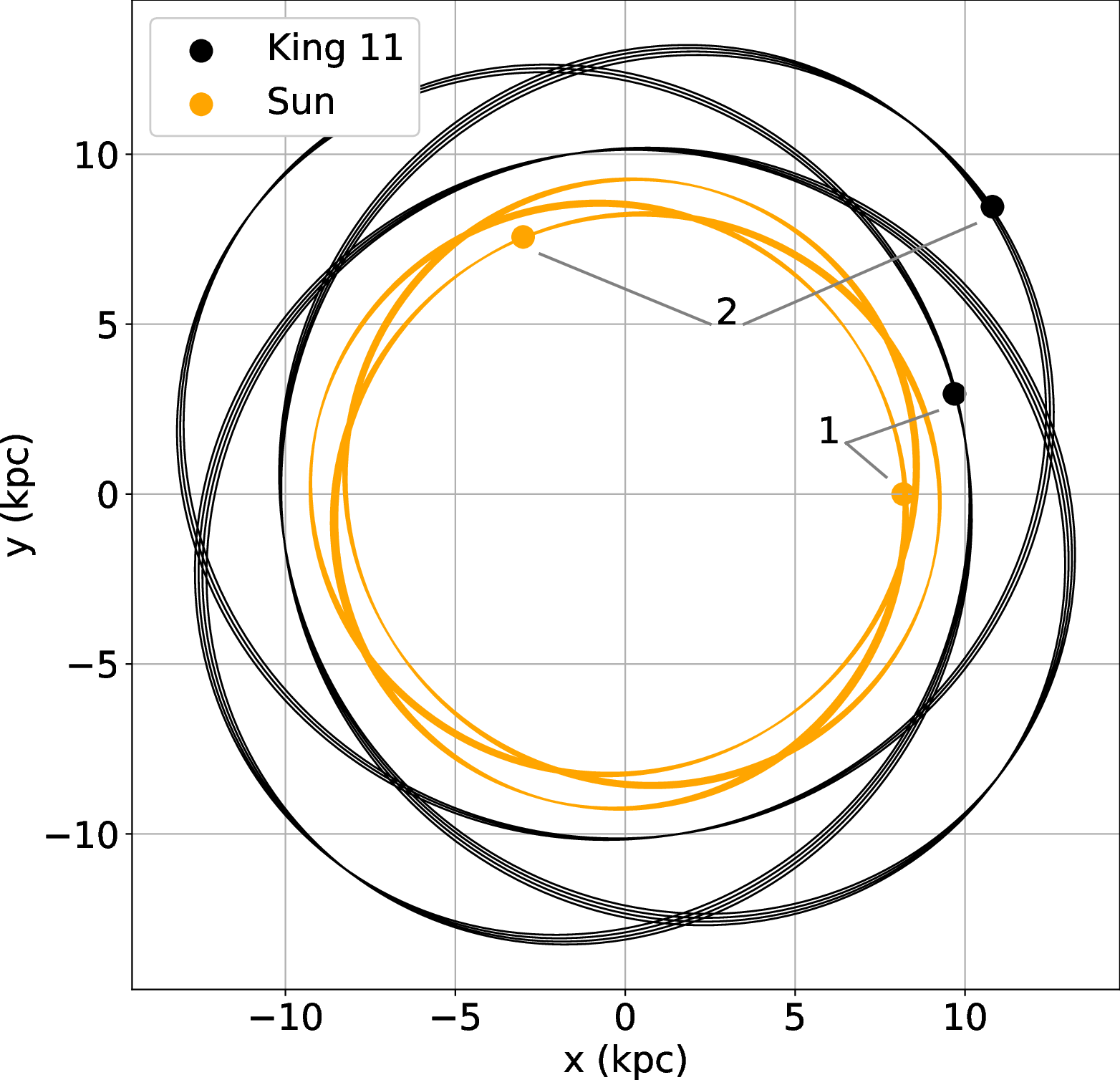}
\caption{
The orbits of King~11 and the Sun in 3D (left panel) 
and in the $XY$ plane (right panel). 
In the left panel, we see the oscillating orbits of the cluster (black), 
and the Sun (orange). 
The red dots show the orbital positions of King 11 and the Sun 
at the time of their approach to a distance of 1.58~kpc, 
which, according to our calculations, happened 0.76 billion years ago.
In the right panel, the orbital positions of the cluster 
and the Sun are given at different times.
Here, the position `1' denotes the current position of the cluster 
($Z_{0}= 0.401\pm 0.010$ kpc) ,
where the cluster's distance from the Sun is $d_{Sun}=3.33$~kpc.
The tentative location of the cluster's birth is shown by the notation `2'.
This corresponds to an epoch of 3.63 Gyr back in time. 
Here, the cluster's distance from the Sun is $d_{Sun}=12.44\pm4.38$~kpc.
But, as explained in Section~\ref{orbit}, the position of the 
cluster's birthplace could be between $Z=-0.320$~kpc and $Z=0.419$~kpc.
}
\label{fig_orbit}
\end{figure}
%%%%%%%%%%%%%%%%%%%%%%%%%%%%%%%%%%%%%%%%%%%%%%%%%%%%%%%%%%%%%%%%%%%%%%%%%%%

%%%%%%%%%%%%%%%%%%%%%%%%%%%%%%%%%%%%%%%%%%%%%%%%%%%%%%%%%%%%%%%%%%%%%%%%%%%
\begin{figure}
%\centering
\includegraphics[height=7.5cm, width=7.5cm]{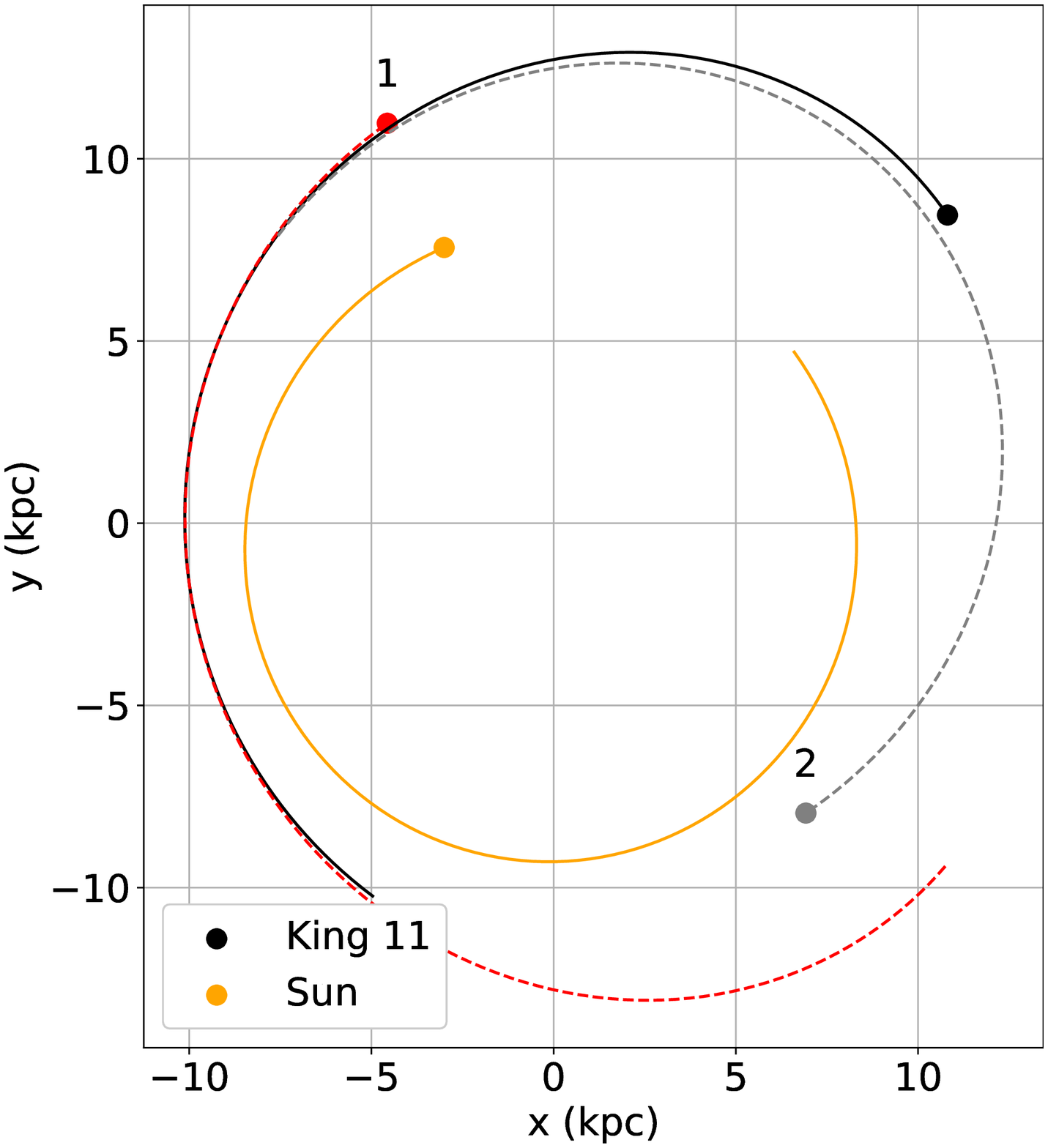}
\includegraphics[height=8.5cm, width=8.5cm]{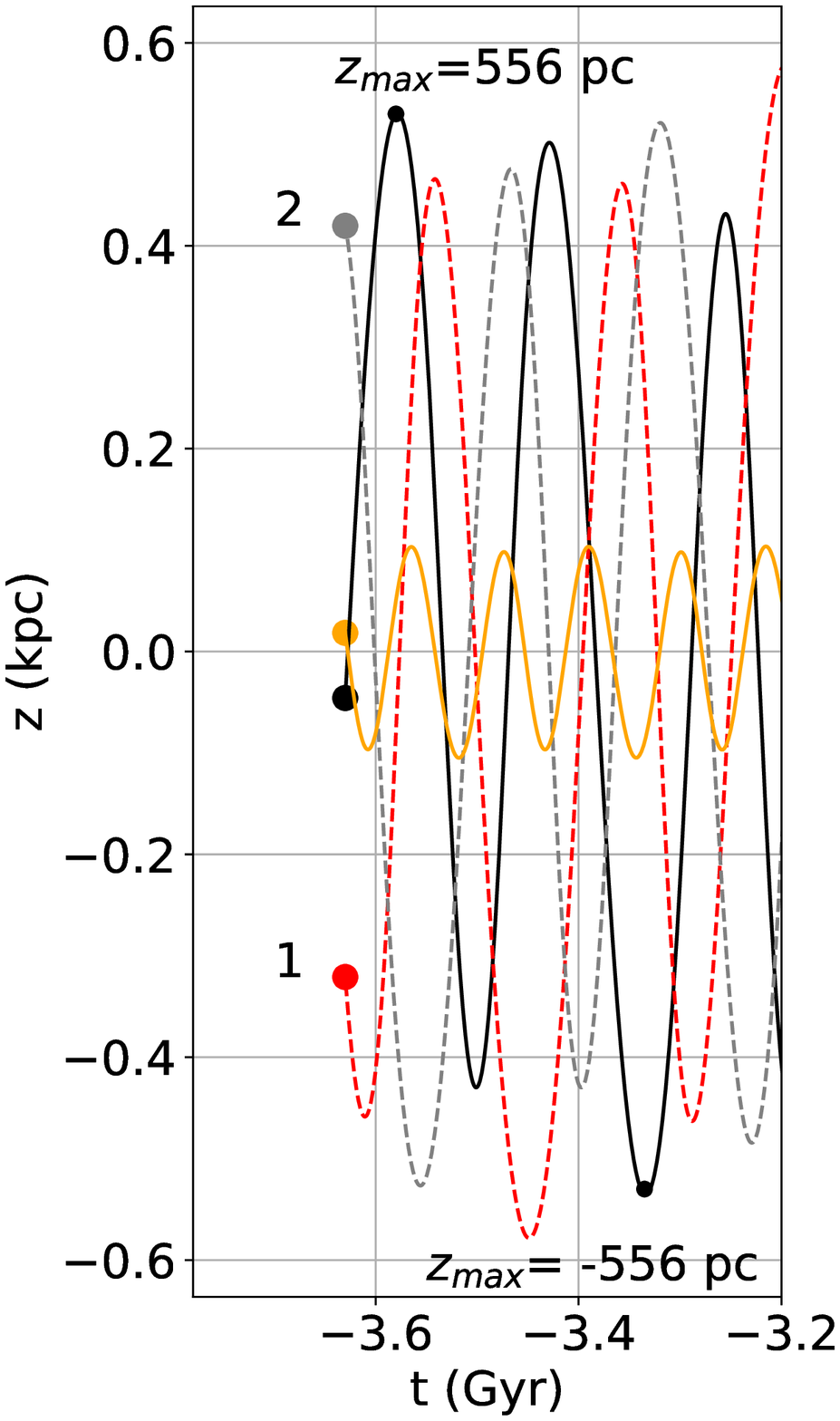}
\caption{
The left panel shows the final segments of the orbits and 
positions of the King~11 cluster (black) and the Sun (orange). 
The calculations are done in the past up to the formation 
epoch of the cluster at $3.63\pm0.42$ Gyr ago in the $XY$ plane of the Galaxy. 
The black dot is the final cluster position according to 
the results of calculations shown in Fig.~\ref{fig_orbit}. 
The gray and red dashed curves show the orbit 
when the errors in PMs, radial velocity, and distance are taken into account, 
where the gray and red dots show the position of the cluster’s birthplace.
The right panel shows the dependence of $Z$-coordinate
of the orbital position on time in past epochs. 
Here notation `1' shows the curve when the errors are added to the original 
input values and we get a $Z_{max}$= $-$0.320 kpc in this case.
Similarly, the notation `2' is used for the case when the errors are 
subtracted. In this case, the value of $Z_{max}$ is 0.419 kpc.
}
\label{fig_orbitXYZ}
\end{figure}
%%%%%%%%%%%%%%%%%%%%%%%%%%%%%%%%%%%%%%%%%%%%%%%%%%%%%%%%%%%%%%%%%%%%%%%%%%%

%%%%%%%%%%%%%%%%%%%%%%%%%%%%%%%%%%%%%%%%%%%%%%%%%%%%%%%%%%%%%%%%%%%%%%%%%%%
\begin{table*}
{\footnotesize
\tabcolsep=0.1cm
\caption{
The input Parameters for the orbit integration of King 11.
}
\vspace{0.5cm}
\centering
\begin{tabular}{cccccc}
\hline\hline
\noalign{\smallskip}
$\mu_\alpha\cos\delta$&$\mu_\delta$&d&$\alpha$ (J2000) & $\delta$ (J2000) & $V_r$\\
mas yr$^{-1}$& mas yr $^{-1}$& kpc & deg & deg & km s$^{-1}$\\
\hline
\noalign{\smallskip}
$-3.391\pm0.006$&$-0.660\pm0.004$ &$3.33\pm0.15$&356.912&68.636&$-24.96\pm5.00$ \\
\hline
\label{tab_orbitin}
\end{tabular}
}
\end{table*}
%%%%%%%%%%%%%%%%%%%%%%%%%%%%%%%%%%%%%%%%%%%%%%%%%%%%%%%%%%%%%%%%%%%%%%%%%%%

%%%%%%%%%%%%%%%%%%%%%%%%%%%%%%%%%%%%%%%%%%%%%%%%%%%%%%%%%%%%%%%%%%%%%%%%%%%
\begin{table*}
\tabcolsep=0.1cm
\caption{
The main orbit parameters resulted from the orbit integration. 
}
\vspace{0.5cm}
\centering
\begin{tabular}{lr}
\hline\hline
\noalign{\smallskip}
Parameter & Value with error\\
\hline
\noalign{\smallskip}
      Apocenter radius (kpc)& $13.80 \pm 0.49$\\
     Pericenter radius (kpc)& $10.10 \pm 0.01$\\
                        $e$ & $0.15  \pm 0.02$\\
	    $Z_{max}$ (kpc) & $0.556 \pm 0.022$\\
	  $Z_{0}$ (kpc)     & $0.401 \pm 0.010$\\
           $U$ (km s$^{-1}$)& $59.98 \pm 2.36$\\
           $V$ (km s$^{-1}$)& $2.50  \pm 4.37$\\
           $W$ (km s$^{-1}$)& $0.24  \pm 0.6$\\
 V$_{spatial}$ (km s$^{-1}$)& $60.20 \pm 2.16$\\
                 $T_r$ (Gyr)& $0.250 \pm 0.006$\\
\hline
\label{tab_orbitout}
\end{tabular}
\end{table*}
%%%%%%%%%%%%%%%%%%%%%%%%%%%%%%%%%%%%%%%%%%%%%%%%%%%%%%%%%%%%%%%%%%%%%%%%%%%

To determine the Galactic orbit of King 11, 
we used galpy package (Bovy 2015) 
which is based on the Python programming language.
We performed orbital integrations using ``MWPotential2014'',
the default galpy potential of the Milky Way (Bovy 2015).
In the set up we used, the Milky Way is represented by 
a three-component model, 
including a halo (radius=16 kpc), a disk and 
an ellipsoidal bulge (size 3$\times$0.28 kpc).
The bulge and disk are described according to 
Miyamoto-Nagai (Miyamoto \& Nagai 1975) expressions.
The spherically symmetric spatial distribution 
of the dark matter density in the halo is described by 
the Navarro Frank-White profile (Navarro et al. 1995).
The Sun's Galactocentric distance is taken as $R_0=8.178$~kpc (McMillan 2017)
and orbital velocity of the Sun is taken as 
$V_0=232.8$~km/s (Gravity Collaboration 2019).

The input data for integrating the cluster's  orbit 
are given in Table~\ref{tab_orbitin}.
As input, we provide the following parameters: 
the mean PM ($\mu_\alpha\cos\delta, \mu_\delta$) 
calculated in Section~\ref{MP},
distance from the Sun (d) from Section~\ref{ISOCHRONE}, 
the cluster's central coordinates ($\alpha$, $\delta$)
taken from Cantat-Gaudin et al. (2018) catalogue
and the radial velocity $V_r$.
The value of $V_r$ was calculated by taking average
of the radial velocity values of the 8 stars we had used in apex determination.

The integration was carried out in the past epoch, 
in a time interval equal to the age (3.63 Gyr) of the cluster,
leading to its possible birthplace. 
Figure~\ref{fig_orbit} shows the 3-dimensional and $XY$-plane projected 
orbits of King~11 and the Sun.

The orbit determination is probabilistic in its nature 
and the results are approximate, because the errors in 
PMs, radial velocity and distance affect them. 
The influence of errors in the input parameters  
(listed in Table~\ref{tab_orbitin})
on the cluster's orbit and the tentative birthplace 
($t=-3.63$~Gyr ago) is shown in the Figure~\ref{fig_orbitXYZ}.
The figure shows the cluster's orbit at the place of its formation. 
We see that the possible birthplace of the cluster can
vary owing to the errors.
In the plane of the Galaxy, 
the distance between the extreme positions is about 3~kpc along the orbit.
The right panel of Figure~\ref{fig_orbitXYZ} shows that the cluster
could have formed both in the north and south hemispheres of the Galaxy.
Also, the cluster in the $Z$-coordinate reaches about 0.556~kpc.

According to the orbit integration results, 
the cluster oscillates (intersects the Galactic plane) 
about 4 times in its radial period ($T_r$),
rising above the plane of the disk up to $0.556 \pm 0.022$~kpc.
We estimated that the cluster's tentative birth place could be anywhere between
$Z$= $-$0.320 and $Z$= 0.419~kpc 
(when errors in PMs and radial velocity are considered).
The current position of the cluster is at $Z_{0}=0.401 \pm 0.01$~kpc.
The cluster's orbit is located entirely outside the Solar orbit,
closer to the edge of the disk.
We determined that the cluster's closest approach to the Sun 
was $\sim$1.6~kpc which happened $\sim$0.76 Gyr ago.

The output parameters with their errors are listed in Table~\ref{tab_orbitout}.
The following output parameters are resulted from the orbit integration using
uncertainties in PMs and radial velocities:
apocentric and pericentric radii (kpc), 
eccentricity ($e$),
the maximum vertical height $Z_{max}$ (kpc), 
the current value of vertical height $Z_{0}$ (kpc), 
heliocentric Galactic rectangular velocities $(U,V,W)$ (km s$^{-1}$), 
spatial velocity $V_{spatial}$ (km s$^{-1}$, calculated from $U,V,W$), 
and radial period $Tr$ (Gyr).

\section{Conclusions}
\label{CON}

We presented a comprehensive photometric and kinematical study 
of the old open cluster King 11 using the Gaia-EDR3 data.
We have used the most probable cluster members 
for a photometric and kinematical follow-up analysis of the King 11. 
The main consequences of the current investigation are as following:

\begin{itemize}

\item The cluster's limiting radius is estimated as 18.51 arcmin using a radial density profile.\

\item Based on the membership probability estimation of stars, 
we identified 676 most probable cluster members of King 11 
that are inside the cluster's limiting radius. 
The mean PM of the cluster is estimated as 
($-3.391\pm0.006$, $-0.660\pm0.004$)
mas yr$^{-1}$ along the RA and DEC directions, respectively.\

\item A total of 13 most probable member BSS of King 11 were detected.
Their population displays a preference of being located 
towards the central region of the cluster. \

\item Using isochrone fitting, the heliocentric distance of the cluster 
is determined as $3.33\pm0.15$ kpc. 
The cluster's age is determined as $3.63\pm0.42$ Gyr by comparing
the cluster's observational CMD with the theoretical isochrones 
given by Bressan et al. (2012). 
The fitted isochrones have a metallicity of Z$_{metal}$=0.011.\

\item The apex coordinates of King~11 are determined using the AD-chart method.
The estimated cluster’s apex is 
$A=267.84^\circ\pm1.01^\circ$, $D=-27.48^\circ\pm1.03^\circ$.
Further, the apex coordinates were tested using the ($\mu_U$, $\mu_T$) diagram.

\item The orbit of the cluster in the Galactic potential is studied.
The scenarios we considered included the cluster's current position 
as well as the cluster's motion in the past.
The spatial velocity of the cluster is equal to 
$60.2 \pm 2.16$~km~s$^{-1}$ relative to the Sun.

\item Our orbit model shows that the closest distance 
that the cluster approached the Sun is 
1.58~kpc about 0.76~Gyr ago. 
We also provide the tentative birth position of the cluster 
which is affected by errors in the input parameters. \

\item The cluster shows an oscillation of 0.556$\pm$0.022~kpc 
in the $Z$-coordinate. \ 

\end{itemize}

\section*{Acknowledgment}
The authors thank the anonymous referee for 
the useful comments that improved the scientific content 
of the article significantly.
This work is supported by the grant from the 
Ministry of Science and Technology (MOST), Taiwan. 
The grant numbers are MOST 105-2119-M-007 -029 -MY3 
and MOST 106-2112-M-007 -006 -MY3.
The reported study was funded by RFBR and DFG 
according to the research project No 20-52-12009.
This work are partly supported by the Russian Foundation for Basic Research (RFBR)
and DFG (grant number 20-52-12009).
This work has made use of data from the European Space Agency (ESA) mission
Gaia (http://www.cosmos.esa.int/gaia), processed by the Gaia Data
Processing and Analysis Consortium (DPAC, http://www.cosmos.
esa.int/web/gaia/dpac/ consortium). Funding for the DPAC has been
provided by national institutions, in particular the institutions 
participating in the Gaia Multilateral Agreement.
In this work we used a library for the Galactic dynamic (galpy) in python created by Bovy 2015.
We are grateful for the useful advice of J. Bovy from
the Department of Astronomy and Astrophysics of the University of Toronto,
in particular, about using the galpy package.

%%%%%%%%%%%%%%%%%%%%%%%%%%%%%%%%%%%%%%%%%%%%%%%%%%%%%%%%%%%%%%%%%%%%%%%%%%%

\end{document}